\documentclass[12pt,epsf]{article}
\usepackage{amsthm}
\usepackage{epsfig, psfrag, graphicx, lscape}
\def\b{\bigbreak}
                                                                                                                                     
\newcommand{\eqnzero}{\setcounter{equation}{0}}
\newcommand{\addeqn}{\addtocounter{section}{1}}

\newcommand{\bay}{\begin{eqnarray}}
\newcommand{\eay}{\end{eqnarray}}
\newcommand{\nopageno}{\thispagestyle{empty}}

\usepackage[figuresright]{rotating}
\usepackage{natbib}
\usepackage{amsmath}
\usepackage{amsfonts}
\usepackage{amssymb}
\usepackage{graphicx}
\usepackage{subfigure}
\usepackage{psfrag}
\usepackage{setspace}
\usepackage{xtab}
\usepackage{verbatim}
\usepackage{chngpage}
\usepackage{lscape}
\usepackage{times}
\usepackage{color}
\begin{document}

\baselineskip=23.3pt
       
\nopageno                                                                                                                              
\begin{center}
{\Large {\bf Post-randomization Biomarker Effect Modification in an HIV Vaccine Clinical Trial}}\\
\b

{\bf Peter B. Gilbert$^{1,*}$, Bryan S. Blette$^{2}$, Bryan E. Shepherd$^{3}$, Michael G. Hudgens$^{2}$}\\

\medskip
{$^1$ Department of Biostatistics, University of Washington and Vaccine and Infectious Disease and Public Health Sciences Divisions, Fred Hutchinson Cancer Research Center, Seattle, Washington, 98109, U.S.A.}\\
{$^2$Department of Biostatistics, University of North Carolina, Chapel Hill, North Carolina, 27599, U.S.A.}\\
{$^3$Department of Biostatistics, Vanderbilt University School of Medicine, Nashville, Tennessee, 37232, U.S.A.}\\ 
$^*${\it e-mail:} pgilbert@scharp.org \\

\vspace{.25in}
Technical Report, Fred Hutchinson Cancer Research Center
May 23, 2018
\end{center}

\noindent \textsc{Summary.}
While the HVTN 505 trial showed no overall efficacy of the tested vaccine to prevent HIV infection over placebo, previous studies, biological theories, and the finding that immune response markers strongly correlated with infection in vaccine recipients generated the hypothesis that a qualitative interaction occurred.  This hypothesis can be assessed with statistical methods for studying treatment effect modification by an intermediate response variable (i.e., principal stratification effect modification (PSEM) methods).  However, available PSEM methods make untestable structural risk assumptions, such that assumption-lean versions of PSEM methods are needed in order to surpass the high bar of evidence to demonstrate a qualitative interaction. Fortunately, the survivor average causal effect (SACE) literature is replete with assumption-lean methods that can be readily adapted to the PSEM application for the special case of a binary intermediate response variable. We map this adaptation, opening up a host of new PSEM methods for a binary intermediate variable measured via two-phase sampling, for a dichotomous or failure time final outcome and including or excluding the SACE monotonicity assumption.  The new methods support that the vaccine partially protected vaccine recipients with a high polyfunctional CD8+ T cell response, an important new insight for the HIV vaccine field.
\\
\noindent {\textsc{Key words}: HIV vaccine; Principal stratification; Randomized clinical trials; Truncation by death; Two-phase sampling.

\vspace{.14in}
\noindent {\bf 1. Introduction}
\eqnzero
\addeqn

The HIV-1 pandemic continues to wreak morbidity and mortality in the U.S. and globally, and a preventive vaccine would be the most effective biomedical tool to end the pandemic.  The most recent efficacy trial of a candidate HIV-1 vaccine -- HVTN 505 -- randomized 2,496 HIV-1 negative volunteers in 1:1 allocation to the DNA/rAd5 vaccine regimen or placebo (Hammer et al., 2013).\nocite{Hammeretal2013short2}  Inoculations were administered at months 0, 1, 3, 6, and the primary objective compared the rate of HIV-1 infection from 6.5 to 24 months between the randomized treatment arms.
The estimated cumulative incidence over this period was 4.62\% (3.15\%) in the vaccine (placebo) group, with cumulative incidence ratio 1.46 (95\% CI 0.82 to 2.63, Wald test $p=0.20$). 
Through a 2-phase sampling design, Janes et al. (2017)\nocite{Janesetal2017short}
studied HIV-1 Envelope-specific CD8+ T cell responses measured by intracellular cytokine staining
2 weeks after the last vaccination (Month 6.5) as a correlate of subsequent HIV-1 infection through 24 months, and found in vaccine recipients a strong inverse correlation between CD8+ T cell polyfunctionality score (PFS) and HIV-1 infection ($p < 0.001$).
Fong et al. (2018)\nocite{Fongetal2018short} followed up this analysis by
 studying Envelope-specific IgG antibodies at the same study visit, and found
that antibodies were also inversely correlated with infection ($p < 0.01$).  Moreover,
Fong et al. found a significant interaction, where having both low antibodies and low PFS was associated with especially high risk.

The surprising strength of these correlates of risk [e.g., estimated cumulative risk 0.160 and 0.034 for vaccinated subgroups with PFS below and above the median response, respectively, which was above and below the estimated cumulative risk of 0.070 for placebo recipients (Figure 4 of Janes et al.)] suggests the possibility that a qualitative interaction occurred, where some marker-defined vaccinated subgroups received partial protection from vaccination whereas others had their risk increased by vaccination.  In addition, two previous efficacy trials of a candidate HIV-1 vaccine that included the same type of vaccine component (rAd5) showed increased risk of infection by vaccination among participants with a specific baseline biomarker value, based on significant interaction tests (Huang et al., 2015 and references therein).\nocite{Huangetal2015}  Moreover, the identified PFS and antibody marker correlates of risk are components of known or putative mechanisms of vaccine protection for several licensed vaccines (Plotkin, 2008), suggesting biological plausibility of a possible qualitative interaction in HVTN 505 based on these markers.

The observations in Janes et al. and Fong et al. of intermediate risk of the placebo group between that of the marker response-defined vaccinated subgroups does not constitute credible evidence of causal vaccine vs. placebo effect modification across the marker response subgroups, because post-randomization selection bias could occur (e.g., Frangakis and Rubin, 2002).
For example, if a genetic factor correlates with the ability of the vaccine to make a high PFS,
and this genetic factor is also prognostic for HIV infection, then 
the observed differences may not reflect a vaccine vs. placebo causal effect.  A proper assessment that does directly assess a causal vaccine effect (eliminating possible selection bias) would compare risk for each marker response vaccinated subgroup to that of the placebo recipient subgroup who would have had the same marker response value if assigned vaccination, which essentially repeats the primary analysis of vaccine efficacy (which is valid based on the randomization) across the marker-defined subgroups -- such analysis may be referred to as ``principal stratification effect modification (PSEM) analysis" or ``principal surrogate analysis" (Frangakis and Rubin, 2002).  Other causal investigations are of equal interest, 
such as assessment of the markers as mediators of vaccine efficacy or as surrogate endpoints for HIV infection.  These analyses tackle distinct questions from effect modification -- namely studying the markers as mechanisms of efficacy and as valid replacement endpoints, respectively (e.g., vanderWeele (2015)\nocite{vanderWeele2015} and Gilbert et al. (2015)) -- and our scope is solely the study of effect modification, which is appropriately tackled by PSEM methods.  A common criticism of PSEM methods is that the subgroups for inference depend on values of both intermediate response endpoints under each treatment assignment, such that no participants have known membership in the subgroups for inference, precluding public health actions/decision-making for subgroups (e.g., Joffe, 2011).\nocite{Joffe2011} 
However, this problem is ameliorated for HVTN 505, because only marker response values if assigned to vaccine are needed, such that PSEM inferences are interpretable in terms of how vaccine efficacy varies over observable subgroups of vaccine recipients.  The interpretation of these inferences is essentially the same as those for the very common objective in randomized trials to assess how treatment efficacy varies over baseline subgroups.

However, available PSEM methods make strong structural risk assumptions that do not hold in our study, and/or require study design augmentations such as close-out placebo vaccination to measure biomarker response (Follmann, 2006) that were not utilized in HVTN 505.  Given the high bar to convince the scientific community of a qualitative interaction, application of these methods would be inadequate -- rather application of assumption-lean PSEM methods that avoid strong untestable assumptions are needed.  
Fortunately, the literature on survivor average causal effect (SACE) methods (e.g., Robins, 1986) is replete with assumption-lean methods that can be readily adapted to the PSEM application for the special case of a binary intermediate response variable.  We map this adaptation, including accommodation of the SACE methods to handle the sub-sampling design of the binary response variable (e.g., two-phase sampling in HVTN 505), which opens up a host of new methods for evaluating a binary intermediate response effect modifier.  While the PFS and antibody markers have underlying mixed binary and continuous distributions, investigation of binary variables based on these markers is important because regulatory agencies often require thresholds of response for vaccine licensure or for bridging of vaccine efficacy to new settings (e.g., Plotkin, 2008).\nocite{Plotkin2008}  The remainder of this article is organized as follows. Section 2 summarizes the relevant SACE and PSEM statistical methods literature and their relationship.  Section 3 defines the PSEM target parameters of interest and their mapping to SACE estimands, and describes identifiability assumptions.  Section 4 summarizes how existing SACE methods of inference can be adapated. 
Section 5 evaluates the new PSEM methods in a simulation study, and   
Section 6 applies these methods to HVTN 505, yielding new insights.  Section 7 concludes with discussion.

\vspace{.14in}

\noindent {\bf 2. SACE and PSEM Problem Statements and their Connection}

{\bf 2.1. SACE Methods}.

A common objective of randomized clinical trials is to assess the effect of a binary treatment $Z$ on an outcome of interest $Y$, where $Y$ is only defined or observable for subjects with a post-randomization intermediate response variable $S$ equal to a certain level (say $S=1$).
Because the intermediate response occurs after randomization, a naive comparison of $Y$ between treatment arms conditional on $S=1$ could be misleading.  A causal estimand of interest
is the SACE.
To formally define the SACE, for each $z=0,1$, let $S(z)$ and $Y(z)$ be potential outcomes of $S$ and $Y$ if a subject is assigned to treatment $Z=z$, where $S(z)=S$ and $Y(z)=Y$ if $Z=z$ is assigned and $S(z)$ and $Y(z)$ are unobserved counterfactuals if $Z=1-z$ is assigned.
Then the SACE is defined as
\begin{eqnarray}
{\rm SACE} \equiv E[Y(1)|S(1)=S(0)=1] - E[Y(0)|S(1)=S(0)=1].  \nonumber
\end{eqnarray}

The large SACE methods literature includes techniques for nonparametric bounds  
\citep{HudgensHoeringSelf2003,Imai2008},
and techniques for sensitivity analysis that estimate the SACE under a spectrum
of selection bias models 
\citep{GilbertBoschHudgens2003,Jemiaietal2007,ShepherdGilbertDupont2011,ChibavanderWeele2011,TchetgenTchetgen2014}, 
with output often including ignorance intervals and estimated uncertainty intervals about the SACE\citep{Vansteelandtetal2006,Richardsonetal2014}.    
Several of these methods make the SACE monotonicity assumption [$P(S(0)\le S(1)) = 1$] and several relax this assumption.

{\bf 2.2. PSEM Methods}.

Another common objective of randomized clinical trials is to assess biomarkers as principal surrogate endpoints for the clinical endpoint of interest. This literature was sparked by Frangakis and Rubin (2002)\nocite{FrangakisRubin2002}, and, with $S^*$ denoting an intermediate response endpoint measured at some fixed time point $\tau$ post-randomization and $Y$ denoting the binary clinical endpoint of interest measured after $\tau$, involves inference about the ``principal effects" or ``causal effect predictiveness" surface \citep{GilbertHudgens2008}:
$CEP(s_1,s_0) = h(risk_1(s_1,s_0),risk_0(s_1,s_0))$,
where
$risk_z(s_1,s_0) \equiv P(Y(z)=1|S^*(1)=s_1,S^*(0)=s_0,Y^{\tau}(1)=Y^{\tau}(0)=0)$
for $z=0,1$.
Here $h(x,y)$ is a contrast function satisfying $h(x,y)=0$ if and only if $x=y$ and $h(x,y) < 0$ for $x < y$, and $Y^{\tau}(z)$ is the indicator that the clinical endpoint occurs by time $\tau$.  The principal surrogate evaluation literature includes frequentist methods based on estimated maximum likelihood 
\citep{Follmann2006,GilbertHudgens2008,GabrielSachsGilbert2015} or on pseudo-score estimating equations \citep{HuangGilbertWolfson2013}, and Bayesian methods \citep{LiTaylorElliott2010}.
Because the principal stratification framework is designed for
studying effect modification and not for determining a valid replacement (i.e. surrogate) endpoint (Gilbert et al., 2015),\nocite{GilbertGabrielHuangChan2015} henceforth we refer to the ``principal surrogate" problem as the more general ``principal stratification effect modification" (PSEM) problem.

{\bf 2.3. Connection between SACE and PSEM Methods}.

Another common objective of randomized clinical trials is to assess biomarkers as principal surrogate endpoints for the clinical endpoint of interest. This literature was sparked by Frangakis and Rubin (2002)\nocite{FrangakisRubin2002}, and, with $S^*$ denoting an intermediate response endpoint measured at some fixed time point $\tau$ post-randomization and $Y$ denoting the binary clinical endpoint of interest measured after $\tau$, involves inference about the ``principal effects" or ``causal effect predictiveness" surface \citep{GilbertHudgens2008}:
Interestingly, despite the close relationship between the two problems, the SACE and PSEM literatures have developed separately.  One reason for this is that for the PSEM problem, $S^*$ is typically categorical, continuous, or censored continuous, thus entailing inference across a spectrum of principal stratum subgroups, whereas the SACE problem makes inference for a single principal stratum. Focusing on a single stratum
has facilitated development of nonparametric and semiparametric SACE methods that need only deal with one or a few terms that are not identified from the observed data, whereas
an attempt to apply SACE methods for each principal stratum defined by a general categorical $S^*$ would face a much larger number
of non-identified terms that grows
with the number of categories.
However, there is an important special case for which the existing SACE methods can be practically adapted-- when $S^*$ is binary 
(i.e., $S^*=1$ versus $S^*=0$ denotes positive versus negative or ``high" versus ``low" response)-- for which only a few principal strata are of interest.

We consider methods assuming either ``equal early clinical risk" (EECR) or ``early no-harm monotonicity" (ENHM), where EECR specifies no individual clinical treatment effects by $\tau$ (i.e., $P(Y^{\tau}(1)=Y^{\tau}(0))=1$) and ENHM specifies no harmful treatment effects by $\tau$ (i.e., $P(Y^{\tau}(1) \le Y^{\tau}(0))=1$).  
For methods assuming ENHM, we focus on the 
special case that $Z=0$ is a control condition such as placebo, and there is no variability of $S^*$ in subjects assigned to $Z=0$, i.e., $P(S^*(0) = 0|Y^{\tau}(0)=0) = 1$.  This ``Constant Biomarker (CB)" case occurs in placebo-controlled 
preventive vaccine efficacy trials that only enroll subjects not previously infected with the pathogen under study and for which the intermediate response endpoint is a readout from a 
validated bioassay designed explicitly to only detect an immune response {\it specific} to the pathogen under study \citep{GilbertHudgens2008}; HVTN 505 is a typical example.  
For methods assuming EECR, we consider both the special Case CB and the general case where $S^*(0)$ varies and a monotonicity assumption holds [$P(S^*(0) \le S^*(1)|Y^{\tau}(1)=Y^{\tau}(0)=0) = 1$], which states that among participants who will not develop a clinical event by $\tau$ regardless of treatment assignment, none have a higher biomarker if assigned $Z=0$ than if assigned $Z=1$.  These scenarios are chosen because they commonly occur in practice and identifiability is relatively easy to achieve.

Under each of the three assumptions [{\bf A:} EECR, $S^*(0)$ varies], [{\bf B:} EECR, Case CB], and [{\bf C:} ENHM, Case CB], we show how to apply previously developed SACE methods to the PSEM problem, which for {\bf A} amounts to inference about $CEP(0,0)$, $CEP(1,0)$, and $CEP(1,1)$, and for
${\bf B}$ and ${\bf C}$ amounts to inference about $CEP(0,0)$ and $CEP(1,0)$.  
Additional original contributions of this work for the PSEM literature include: 
(i) to allow relaxing the strong EECR assumption-- which has been made for almost all published methods-- to ENHM; and
(ii) to provide methods that can be straightforwardly applied for the common situation in practice where no adequate baseline predictors of $S^*$ are available and closeout placebo vaccination was not performed [previous frequentist methods, such as
Follmann (2006)\nocite{Follmann2006} and subsequent work, require at least one of these design augmentations].
Like (i), contribution (ii) largely broadens the set of applications 
to which PSEM analysis can be done.
%
A third novel development addresses the issue that most SACE methods do not explicitly accommodate sub-sampling designs (e.g., case-control/two-phase or case-cohort) for measuring $S$ [Shepherd, Redman, and Ankerst (2009) \nocite{ShepherdRedmanAnkerst2009} is one exception, which we apply here], whereas most PSEM evaluation methods do.   Because sub-sampling designs are normative in PSEM applications, we focus on this setting, with implication that some of the previously developed SACE methods must be extended.
For simplicity we focus on an adjustment based on inverse probability weighting, but more efficient methods would be straightforward to implement.

\vspace{.14in}

\noindent {\bf 3. Notation, Target Parameters, Identifiability Assumptions}

{\bf 3.1. Additional Notation}.

Let $W$ be a vector of baseline covariates measured in everyone, and $R$ be the indicator of whether the binary endpoint $S^*$ is measured at $\tau$.  We allow $S^*$ to be missing due either to design 
and/or to happenstance reasons outside of the control of the investigator. To fit the most common applications in clinical trials we assume $Y$ is binary; in the case of a time-to-event outcome, $Y=I(T \le t)$, where $T$ is the time from randomization until the endpoint subject to right-censoring and $t > \tau$ is a fixed time point of interest.  
Note that $S^*$ is undefined if $T \le \tau$, which we denote as $S^* = *$.  


{\bf 3.2. Target Parameters for Principal Stratification Effect Modification}.

Our goal is inference about $CEP(0,0)$, $CEP(1,0)$, and $CEP(1,1)$, where the last parameter is only relevant in scenario {\bf A} where $S^*(0)$ is non-constant. Three criteria for a useful intermediate response endpoint that have been discussed include average causal necessity (ACN) and average causal sufficiency (ACS) (Gilbert and Hudgens, 2008), as well as 
wide variability of $CEP(s_1,s_0)$ across marker values (effect modification).
For scenarios {\bf B} and {\bf C}
these conditions amount to
$CEP(0,0) = 0$ (ACN), $CEP(1,0) \neq 0$ (ACS), and 
$CEP(1,0)$ largely different from $CEP(0,0)$. Under Case CB and assumptions A1--A4 defined in the next section, ACN and ACS both hold if and only if the binary intermediate endpoint satisfies the Prentice (1989)\nocite{Prentice1989} definition of a valid surrogate/replacement endpoint (Gilbert et al., 2015).  Because a valid Prentice surrogate is very useful in practice, checking this definition is a useful application of the methods.

Define overall risks 
$risk_z \equiv P(Y(z)=1|Y^{\tau}(1)=Y^{\tau}(0)=0)$ for $z=0,1$, and note
\begin{eqnarray}
risk_z = p(0,0) risk_z(0,0) + p(1,0) risk_z(1,0) + p(1,1)risk_z(1,1) \hspace{.2in} {\rm for} \hspace{.1in} z=0,1, \label{eq: ben}
\end{eqnarray} 

\noindent where $p(s_1,s_0) \equiv P(S^*(1)=s_1,S^*(0)=s_0|Y^{\tau}(1)=Y^{\tau}(0)=0)$ for
$(s_1,s_0) \in \lbrace (0,0),$ \newline $(1,0),
 (1,1) \rbrace$. [Equation (\ref{eq: ben}) uses the fact that $p(0,1)=0$ under the 
assumptions we use.] These $p(\cdot, \cdot)$'s measure the prevalence of intermediate response subgroups in the early always survivors (EAS) principal stratum defined by
$\lbrace Y^{\tau}(1) = Y^{\tau}(0) = 0 \rbrace$.  

The $risk_z$ and $risk_z(s_1,s_0)$ parameters measure risks in subsets of the EAS principal stratum; for scenario {\bf C} we will also consider parameters measuring risks in subsets of the 
 `early protected' (EP) principal stratum defined by $\lbrace Y^{\tau}(1)=0,Y^{\tau}(0)=1 \rbrace$:
 $risk^{EP}_z(s_1,s_0)$ \newline $ \equiv P(Y(z)=1|S^*(1)=s_1,S^*(0)=s_0,Y^{\tau}(1)=0,Y^{\tau}(0)=1)$ and the marginal risks for $z=1$ defined as
$mrisk_1 \equiv P(Y(1)=1|Y^{\tau}(1)=0)$ and $mrisk_1(s_1) \equiv P(Y(1)=1|Y^{\tau}(1)=0,S^*(1)=s_1)$ for $s_1=0,1$.

The observation motivating this work is 
that a contrast in $risk_1$ and $risk_0$ is a standard SACE (with intermediate event $S = 1-Y^{\tau}$), a contrast in $risk_1(0,0)$ and $risk_0(0,0)$ is a standard SACE (with intermediate event $S=[1-Y^{\tau}][1-S^*]$), 
a contrast in $risk_1(1,1)$ and $risk_0(1,1)$ is a standard SACE (with intermediate event $S=[1-Y^{\tau}]S^*$), 
and, whereas a contrast in $risk_1(1,0)$ and $risk_0(1,0)$ is not a standard 
SACE, these two parameters are identified from (\ref{eq: ben}) and the other parameters.
Thus any two existing SACE methods can be employed to estimate the two sets of
means/probabilities $\lbrace risk_1, risk_0 \rbrace$ and  $\lbrace risk_1(0,0),$ \newline $risk_0(0,0) \rbrace$, and another for $\lbrace risk_1(1,1),risk_0(1,1) \rbrace$ in scenario {\bf A}, and then equation (\ref{eq: ben}) is applied to yield estimates of the remaining two probabilities $\lbrace risk_1(1,0), risk_0(1,0)\rbrace $ via
\begin{eqnarray}
risk_z(1,0) = \left[ risk_z - p(0,0) risk_z(0,0) - p(1,1) risk_z(1,1) \right] \slash 
p(1,0). \label{eq: riskz10}
\end{eqnarray}

The case of an additive contrast function $h(x,y) = x-y$ represents the traditional SACE estimands, for which we have 
%
\begin{eqnarray}
CEP(0,0) & = & {\rm SACE}(0,0) = risk_1(0,0) - risk_0(0,0) \label{eq: CEP00} \\
CEP(1,1) & = & {\rm SACE}(1,1) = risk_1(1,1) - risk_0(1,1) \label{eq: CEP11} \\
CEP(1,0) & = & \frac{1}{p(1,0)}\left[ {\rm SACE}_{mar} - p(0,0) {\rm SACE}(0,0) - p(1,1) {\rm SACE}(1,1) \right], \label{eq: CEP10}
\end{eqnarray}

\noindent where 
${\rm SACE}_{mar} \equiv risk_1 - risk_0$.
However, we develop the results in terms of pairs of risks, in order that they apply for general contrast functions.
In Case CB the above formulas simplify, with $\lbrace risk_1(1,1),risk_0(1,1)\rbrace$ and $CEP(1,1)$ vanishing. 

{\bf 3.3. Assumptions}.

Throughout we make a baseline set of assumptions made in essentially all previous frequentist SACE papers. We assume the $(Z_i,W_i,R_i,Y^{\tau}_i(1),Y^{\tau}_i(0),S^*_i(1),S^*_i(0),Y_i(1),Y_i(0))$, $i=1, \ldots , n,$ are iid [with observed data $O_i = (Z_i,W_i,R_i,R_i S^*_i,Y^{\tau}_i,Y_i)$, with $S^*_i$ only observed if $R_i=1$] and assume 
{\bf A1} SUTVA (Stable Unit Treatment Value Assumption);
{\bf A2} {\it Ignorable Treatment Assignment:}
Conditional on $W$, $Z$ is independent of 
$(Y^{\tau}(1),Y^{\tau}(0),$ \newline $S^*(1),S^*(0),Y(1),Y(0))$; and 
{\bf A3} {\it No Censoring or Random Censoring}: If $Y$ is binary, $Y$ is observed for all subjects; if $Y$ is time-to-event, right-censoring $C(z)$ is random conditional on $W$ ($T(z) \perp C(z)| W$ for $z=0,1$).
As noted above, we develop the methods under each of two assumptions regarding early clinical events before the intermediate endpoint is measured at $\tau$:

\vspace{.1in}
\noindent {\bf A4} {\it Equal Early Clinical Risk (EECR)}: $P(Y^{\tau}(1) =  Y^{\tau}(0)) = 1$ 

\vspace{.1in}
\noindent {\bf A4$^{\prime}$} {\it Early No-Harm Monotonicity (ENHM)}: $P(Y^{\tau}(1) \le Y^{\tau}(0)) = 1$
\vspace{.1in}

\noindent For each of A4 and A4$^{\prime}$ 
we also assume $p(1,0) > 0$, which holds trivially for any PSEM evaluation application of interest.
A4$^{\prime}$ is a standard SACE monotonicity assumption for the analysis of $\lbrace risk_1,risk_0 \rbrace$, which 
weakens the EECR assumption made in \citet{GilbertHudgens2008} and almost all subsequent PSEM evaluation papers.
This weakening is practically important, given that for many applications EECR is not plausible. 
A4 and A4$^{\prime}$ have implications that can be tested by comparing the rates of $Y=1$ by $\tau$ between the two treatment arms.

For scenario {\bf A} where $S(0)$ varies we also reduce the number of non-identified terms via a marker monotonicity assumption:

\vspace{.1in}
\noindent {\bf A5} {\it Marker Monotonicity}: $P(S^*(0) \le S^*(1)|Y^{\tau}(1)=Y^{\tau}(0)=0) = 1$.
\vspace{.1in} 

%
%

%
%

Based on the above stated assumptions, the scenarios under which we develop the methods can be re-stated as
[{\bf A:} A1--A5, $S(0^*)$ varies], [{\bf B:} A1--A4, Case CB], and [{\bf C:} A1--A3, A4$^{\prime}$, Case CB].  

{\bf 3.4. Identifiability of $CEP(s_1,s_0)$}.

We now consider identifiability of $CEP(s_1,s_0)$, highlighting how A4$^{\prime}$, A4, A5, and Case CB reduce the number of non-identified terms.
In this section we describe the number of fixed sensitivity parameters that are needed to achieve nonparametric identifiability, and in the next section give examples of how specific models under two SACE methods provide identifiability.

{\it A1--A3, A5, $S(0)$ varies.}
We start with a slightly weaker assumption set than scenario ${\bf A}$.
Our objective is to estimate the three $CEP(s_1,s_0)$ parameters [for $(s_1,s_0) \in \lbrace (0,0), (1,0), (1,1) \rbrace$],
%
by estimating each of the three risk pairs 
$\lbrace risk_1(0,0),risk_0(0,0)\rbrace$, $\lbrace risk_1(1,1),risk_0(1,1)\rbrace$, and 
$\lbrace risk_1(1,0),risk_0(1,0)\rbrace$.
Because $(Y^{\tau}(1),S^*(1))$ and \newline $(Y^{\tau}(0),S^*(0))$ are never both observed,
these risk pairs
are not identified from A1--A3, A5.  
Following \citet{ShepherdGilbertDupont2011}, without additional assumptions 3 sensitivity parameters are needed to nonparametrically identify each of the three risk pairs, totaling 9 sensitivity parameters.  For example, define $$\pi(s_1,s_0) \equiv P(S^*(1)=s_1,S^*(0)=s_0,Y^{\tau}(1)=Y^{\tau}(0)=0),$$  
$$\pi_1(s_1,s_0) \equiv P(S^*(1) = s_1,Y^{\tau}(1)=0|S^*(0)=s_0,Y^{\tau}(0)=0,Y(0)=1),$$ 
$$\pi_0(s_1,s_0) \equiv P(S^*(0)=s_0,Y^{\tau}(0)=0|S^*(1)=s_1,Y^{\tau}(1)=0,Y(1)=1).$$ 
Then $\lbrace risk_1(0,0), risk_0(0,0)\rbrace$, $\lbrace risk_1(1,1),risk_0(1,1)\rbrace$, and $\lbrace risk_1(1,0),risk_0(1,0)\rbrace$ are identified by the three triplets of parameters $\{\pi(0,0), \pi_1(0,0), \pi_0(0,0)\}$, $\{\pi(1,1), \pi_1(1,1),$ \newline $\pi_0(1,1)\}$, and $\{\pi(1,0), \pi_1(1,0), \pi_0(1,0)\}$, respectively.  From equation (\ref{eq: ben}), each $risk_z$ for $z=0,1$ is also identified by these 9 sensitivity parameters because $p(s_1,s_0)=\pi(s_1,s_0)/$ \newline $\{\pi(0,0)+\pi(1,1)+\pi(1,0)\}$.  Other sensitivity parameterizations can also achieve identifiability, but at least 9 sensitivity parameters will still be required.



{\it Under A1--A3, A4$^{\prime}$, A5, $S(0)$ varies.} Adding A4$^{\prime}$ to A1--A3 and A5,
$\pi(1,1)$ and $risk_0(1,1)$ are nonparametrically identified as $P(S^*(0)=1,Y^{\tau}(0)=0)$ and $P(Y(0)=1|S^*(0)=1,Y^{\tau}(0)=0)$.  Therefore, identifying $risk_1(1,1)$ requires only one sensitivity parameter, $\pi_0(1,1)$.  Identifying $risk_z(0,0)$ for $z=0,1$ requires specifying three parameters, $\{\pi(0,0),\pi_1(0,0), \pi_0(0,0)\}$.  Once $\pi(0,0)$ is specified, $\pi(1,0)$ is identified as $\pi(1,0)=P(S^*(0)=0,Y^{\tau}(0)=0)-\pi(0,0)$; also everyone with $\lbrace S^*(0)=0,Y^{\tau}(0)=0\rbrace $ who does not have $\lbrace S^*(1)=0,Y^{\tau}(1)=0\rbrace$ has to  have $\lbrace S^*(1)=1,Y^{\tau}(1)=0 \rbrace$, and therefore specifying $\pi_1(0,0)$ also fixes $\pi_1(1,0)=1-\pi_1(0,0)$ and thus identifies $risk_0(1,0)$.  Hence, only one additional sensitivity parameter, $\pi_0(1,0)$, is needed to identify $risk_1(1,0)$, bringing the total number of sensitivity parameters to 5.


{\it Under A1--A5, $S(0)$ varies [Scenario {\bf A}].}
Strengthening A4$^{\prime}$ to A4, $\pi(1,1)$ and $risk_0(1,1)$ are still nonparametrically identified, and identification of $risk_1(1,1)$ still requires only one sensitivity parameter, $\pi_0(1,1)$.  This same sensitivity parameter also identifies $risk_1(1,0)$ as $\pi_0(1,0)=1-\pi_0(1,1)$ because by A4 everyone with $(S^*(1)=1,Y^{\tau}(1)=0)$ must have $Y^{\tau}(0)=0$.  Then $\pi(0,0)$ and $risk_1(0,0)$ are also nonparametrically identified.  Therefore, to identify $risk_0(0,0)$ only the sensitivity parameter $\pi_1(0,0)$ needs to be specified.  As before under A4$^{\prime}$, specifying $\pi_1(0,0)$ also identifies $risk_0(1,0)$.  Hence a total of two sensitivity parameters are needed in scenario {\bf A}.


{\it Under A1--A4, Case CB [Scenario {\bf B}].}
Scenario {\bf B} is similar to scenario {\bf A} except that now the cell $(S^*(0)=1,S^*(1)=1,Y^{\tau}(0)=0,Y^{\tau}(1)=0)$ is empty.  This implies that $risk_z(1,1)$ is undefined, for $z=0,1$.  Also, in addition to $risk_1(0,0)$ being identified, $risk_1(1,0)$ is now also nonparametrically identified.  Similar to scenario {\bf A}, $risk_0(0,0)$ is identified with a sensitivity parameter $\pi_1(0,0)$ which then also identifies $risk_0(1,0)$.  Therefore, only one sensitivity parameter is needed for scenario {\bf B}.

%

{\it Under A1--A3, A4$^{\prime}$, Case CB [Scenario {\bf C}].}
Identifiability is more challenging when relaxing A4 to A4$^{\prime}$, as 3 sensitivity parameters (e.g., $\pi(0,0), \pi_0(0,0)$, and $\pi_1(0,0)$) are needed to identify $\lbrace risk_1(0,0),risk_0(0,0)\rbrace$.  The sensitivity parameters $\pi(0,0)$ and $\pi_1(0,0)$ also identify $risk_0(1,0)$.  Moreover, one additional sensitivity parameter, $\pi_0(1,0)$, is required to identify $risk_1(1,0)$.  

{\bf 3.5. Mapping Existing SACE Methods to Estimate Target Parameters}.

As mentioned in Section 2.2, contrasts of the various risk parameters can be estimated using SACE methods by appropriately defining the intermediate endpoint $S$ and applying methods that correspond to the implied monotonicity assumptions in $S$ (written as $P[S(1) \leq S(0)]=1$) based on the specific scenario.  Table 1 summarizes how SACE methods in general can be used to estimate $CEP(s_1,s_0)$ for scenarios {\bf A}, {\bf B}, and {\bf C}.

\begin{table}\small
\caption{Use of SACE Methods [With Numbers of Required Sensitivity Parameters (S.P.s)] for Estimation of $CEP(s_1,s_0)$ for $(s_1,s_0) = (0,0), (1,0), (1,1)$}
\centering
\begin{tabular}{lccc} \hline
   						    & \multicolumn{3}{c}{Assumption Sets} \\ \hline
Target Parameter & [{\bf A:} A1--A5, & [{\bf B:} A1--A4,  & [{\bf C:} A1--A3, A4$^{\prime}$, \\ 
                 &  $S(0)$ varies]    & Case CB]           & Case CB]
\\ \hline
${\rm SACE}_{mar} (risk_1, risk_0)$  & Nonpar.    & Nonpar. & Any SACE  \\
$risk_z \equiv$ & ident.  & ident. & method w/ \\
$P(Y(z)=1|S(1)=S(0)=1)$ &    &      &  monot. A4$^{\prime}$$^a$ \\
w/ $S \equiv 1-Y^{\tau}$     & (0 S.P.s)           & (0 S.P.s)   &  (1 S.P.) \\
			       &	    &			    &              \\
${\rm SACE}(0,0) (risk_1(0,0), risk_0(0,0))$ & Any SACE  &   Any SACE & Any SACE  \\
$risk_z(0,0) \equiv$  & method w/ & method w/ & method w/o \\
$P(Y(z)=1|S(1)=S(0)=1)$  & monot. A5$^{b,c}$  & monot. A5$^c$  & monot.$^d$  \\
w/ $S \equiv [1-Y^{\tau}][1-S^*]$   &  (1 S.P.) & (1 S.P.)  & (3 S.P.s) \\
& &  & \\  
${\rm SACE}(1,1) (risk_1(1,1), risk_0(1,1))$ & Any SACE   &   N/A & N/A \\
$risk_z(1,1) \equiv$ & method w/ & N/A & N/A \\
$P(Y(z)=1|S(1)=S(0)=1)$  & monot. A5$^e$  & N/A & N/A \\
w/ $S \equiv [1-Y^{\tau}]S^*$ & (1 S.P.)  & N/A  & N/A \\
& &			  & \\
Mixing parameters    & Nonpar.    & Nonpar. & Nonpar.  \\
$p(0,0)$, $p(1,0)$, $p(1,1)$     & ident.  & ident. & ident. \\
& (0 S.P.s) & (0 S.P.s) & (0 S.P.s)  \\ 
\hline \\
\end{tabular}
\newline
\noindent
$^a$The monotonicity assumption for the SACE method is $P(S(0) \le S(1)) = 1$ with
$S \equiv 1-Y^{\tau}$. This assumption is A4$^{\prime}$, which holds for all scenarios {\bf A, B, C}. \newline
\noindent $^b$The monotonicity assumption for the SACE method is $P(S(1)\le S(0)) = 1$ with $S \equiv [1-Y^{\tau}][1-S^*]$. \newline
$^c$Because of A4 in scenarios {\bf A, B}, the monotonicity assumption expressed in footnote b simplifies to $P(S^*(0) \le S^*(1)|Y^{\tau}=0) = 1$, which holds by A5 in scenario {\bf A} and by Case CB in scenario {\bf B}. \newline
\noindent $^d$Under A4$^{\prime}$ and Case CB in scenario {\bf C}, the monotonicity assumption expressed in footnote b amounts to the assumption that
no $Z=1$ participants with a negative marker response at $\tau$ 
would be protected by $\tau$. This assumption is often difficult to justify and hence we do not consider methods making this assumption. \newline
\noindent $^e$The monotonicity assumption for the SACE method is $P(S(0)\le S(1)) = 1$ with $S \equiv [1-Y^{\tau}]S^*$. Because of A4 in scenario {\bf A}, the monotonicity assumption 
expressed in footnote d simplifies to $P(S^*(0) \le S^*(1)|Y^{\tau}=0) = 1$, which holds by A5 in scenario {\bf A}. 
\end{table}

Under scenario {\bf A}, SACE$_{mar}$ is nonparametrically identified by A4.  A contrast in \newline $risk_1(0,0)$ and $risk_0(0,0)$ (i.e., $SACE(0,0)$) can be estimated using a standard SACE method under the assumption of monotonicity with intermediate event $S=(1-Y^{\tau})(1-S^*)$.  Similarly, a contrast in $risk_1(1,1)$ and $risk_0(1,1)$ (i.e., $SACE(1,1)$) can also be estimated with a standard SACE method under monotonicity but now with the intermediate event $S=(1-Y^{\tau})S^*$.  The mixing parameters, $p(s_1,s_0)$, are all nonparametrically identified so a contrast in $risk_1(1,0)$ and $risk_0(1,0)$ can be estimated from the other risk parameters and the mixing parameters based on equation (\ref{eq: ben}).  
 
Estimation under scenario {\bf B} is identical to that of scenario {\bf A} except that $risk_1(1,1)$ and $risk_0(1,1)$ vanish (because the principal stratum defined by $\left[ 1-Y^{\tau}(0)\right] S(0)^*= $ \newline $\left[ 1-Y^{\tau}(1)\right] S(1)^*=1$ is empty), simplifying estimation.  Lastly, for scenario {\bf C}, one can estimate SACE$_{mar}$ using a standard SACE method assuming monotonicity with the intermediate event $S=1-Y^{\tau}$.  A contrast in $risk_1(0,0)$ and $risk_0(0,0)$ can be estimated using a standard SACE method that does not assume monotonicity. And, as with scenario {\bf B}, $risk_1(1,1)$ and $risk_0(1,1)$ vanish, thereby simplifying estimation of $risk_1(1,0)$ and $risk_0(1,0)$. 

Standard SACE methods focus inference on a single principal stratum: $CEP(s_1,s_0)$ for a fixed $(s_1,s_0)$.
However, if we make inference on contrasts in the $CEP(s_1,s_0)$, such as $\mu \equiv CEP(1,0) - CEP(0,0)$, then our set-up constrains $\mu$ to values narrower than the maximum possible range -2 to 2.  For example,
for scenario {\bf B}, setting $\beta_0=0$ implies $risk_0(1,0) = risk_0(0,0)$, which leaves each of $CEP(1,0)$ and $CEP(0,0)$ free to vary over the maximum possible range as for any standard SACE method but constrains $\mu$ to -1 to 1.  Thus making inference on contrasts of $CEP(s_1,s_0)$ does not achieve just-nonparametric identifiability as does inference on the individual $CEP(s_1,s_0)$ parameters. 
This should be borne in mind when inference is made on $CEP$ contrasts as well as on the individual parameters.

\vspace{.14in}

\noindent {\bf 4. Estimation of the Target Parameters Using SACE Methods}

SACE methods provide estimators of the two means constituting each SACE.  
For concreteness, for each scenario {\bf A--C} we show how the estimation works using Shepherd, Gilbert, and Dupont's (2011) SACE method, which in scenarios {\bf A, B} simplifies to 
a SACE method described in several papers 
including Gilbert, Bosch, and Hudgens (2003)\nocite{GilbertBoschHudgens2003} (henceforth GBH) and Hudgens and Halloran (2006).\nocite{HudgensHalloran2006}
For simplicity and generality we focus on inverse probability weighting (IPW) extensions of existing SACE methods.  Also for simplicity, we focus on semiparametric efficient estimators given the data $(Z,R,Y^{\tau},S^*,Y)$ not including baseline covariates $W$, which amount to sample means with or without IPW as needed.  Moreover, we focus on the case that $Y$ is binary and not subject to right-censoring; Web Appendix B summarizes how the methods translate to $Y=I(T \le t)$ with $T$ subject to right-censoring before $t$.  We use general estimating function notation so that users preferring to use more efficient estimators leveraging information in $W$ 
[e.g., \citet{ZhangTsiatisDavidian2008,vanderLaanRose2011}] 
may substitute alternative estimating functions into the estimating equations.

{\bf 4.1. General IPW Estimation}.
 
The SACE estimators involve estimation of identified terms
$E[Y|S=1,Z=z]$ for subgroups $S=1$ [with $S=1-Y^{\tau}$, $(1-Y^{\tau})S^*$, or $(1-Y^{\tau})(1-S^*)$] and of terms $E[S^*|Y^{\tau}=0,Z=z]$, where $S^*$ is measured at time 
$\tau$ and is subject to missingness.
Define the probability of observing $S^*$ as
$\pi(O) \equiv P(R=1|O)$, where $O$ is observed data, i.e., ($Z, W, Y^{\tau}, Y$).
We assume $S^*$ is missing at random, $\pi(O) = P(R=1|O,S^*)$, and that
$\pi(O)$ is bounded away from zero,
$\pi(O) \ge \varepsilon$ with probability 1 for some fixed $\varepsilon > 0$.

Following standard IPW estimation, we specify 
a model $\pi(O,\psi)$ for $\pi(O)$ (e.g., logistic regression),
and estimate the unknown parameter $\psi$ by 
maximum likelihood, yielding $\widehat \pi_i =
\pi(O_i,\widehat \psi)$.
Efficiency and robustness may be improved by calibrating the estimated weights $\widehat \pi_i(W_i)$ accounting for $W_i$
 (e.g., Robins, Rotnitzky, and Zhao, 1994;\nocite{RobinsRotnitzkyZhao1994} Rose and van der Laan, 2011; Saegusa and Wellner, 2013\nocite{WellnerSaegusa13}).


{\bf 4.2. Dichotomous Outcome SACE Methods Under Scenario {\bf A}}.

For scenario {\bf A}, the first step
is to estimate the terms that are nonparametrically identified-- \newline
$\lbrace risk_1,risk_0\rbrace$, $p(0,0)$, $p(1,0)$, and $p(1,1)$.  
Each $risk_z$ for $z=0,1$ can be estimated by any preferred method for estimating
a mean, most simply by solving $\sum_{i=1}^n U^{0z}_i(O_i;risk_z)=0$ with
estimating function $U^{0z}(O_i;risk_z) \equiv (1-Y^{\tau}_i)I(Z_i=z)(Y_i - risk_z)$.

Given $p(0,0) = P(S^* = 0|Z=1,Y^{\tau}=0)$, if full data were available, then a simple approach would estimate $p(0,0)$ by solving
$\sum_{i=1}^n U^{01}_i(O_i;p(0,0)) = 0$ with estimating function
$U^{01}(O_i;p(0,0)) \equiv (1-Y^{\tau}_i)Z_i(1-S^*_i - p(0,0))$, 
with convention
$U^{01}(O_i;p(0,0))=0$ if $S^*_i=*$.  
The IPW version of this equation is
$\sum_{i=1}^n R_i U^{01}_i(O_i;p(0,0)) \slash \pi(O_i,\widehat \psi) = 0$.
The parameter $p(1,1)$ may be estimated similarly by solving
$\sum_{i=1}^n R_i U^{00}_i(O_i;p(1,1)) \slash \pi(O_i,\widehat \psi) = 0$
with $U^{00}_i(O_i;p(1,1)) \equiv (1-Y^{\tau}_i)(1-Z_i)(S^*_i - p(1,1))$,
again with convention $U^{00}_i(O_i;p(1,1))=0$ if $S^*_i=*$.
%
%
With $\widehat p(0,0)$ and $\widehat p(1,1)$, we then 
set $\widehat p(1,0) = 1 - \widehat p(0,0) - \widehat p(1,1)$.

Lastly, we estimate each pair $\lbrace risk_1(0,0),risk_0(0,0)\rbrace$ and
$\lbrace risk_1(1,1),risk_0(1,1)\rbrace$ with a SACE method that assumes monotonicity.
For concreteness we summarize how the GBH method accomplishes this task, and then show how GBH easily extends to accommodate IPW. 

{\it Standard Semiparametric SACE Method (Notation as in GBH).} 
Consider the odds ratio selection bias model with user-specified fixed sensitivity parameter
$\beta_0$:

\vspace{.1in}
\noindent {\bf B.0} 
${\rm exp}(\beta_0) = \frac{P(S(1)=1|S(0)=1,Y(0)=1) \slash \left\{ 1 - P(S(1)=1|S(0)=1,Y(0)=1) \right\} }{ P(S(1)=1|S(0)=1,Y(0)=0) \slash \left\{ 1 - P(S(1)=1|S(0)=1,Y(0)=0) \right\} }$.

\vspace{.1in}
\noindent Under A1--A3, {\bf B.0}, monotonicity [$P(S(1)\le S(0))=1$], and positivity [$P(S(1)=0,S(0)=1) > 0$,
$P(S(1)=1,S(0)=1)>0$], the two parameters of interest,
$P^{11}(z) \equiv P(Y(z)=1|S(1)=S(0)=1)$ for $z=0,1$, are nonparametrically identified.
With $w_0(y;\alpha_0,\beta_0) \equiv P(S(1)=1|S(0)=1,Y(0)=y)$ for $y=0,1$, 
these assumptions equivalently specify $w_0(y;\alpha_0,\beta_0) =
\lbrace 1 + {\rm exp}(-\alpha_0 -\beta_0 y) \rbrace^{-1}$ for $y=0,1$ (Jemiai et al., 2007).

By monotonicity $P^{11}(1)=P(Y(1)=1|S(1)=1)$, such that $P^{11}(1)$ is estimated by solving 
$\sum_{i=1}^n U^1(O_i;P(Y(1)=1|S(1)=1)) = 0$ 
where
$U^1(O_i;P(Y(1)=1|S(1)=1)) = Z_iS_i(Y_i - P(Y(1)=1|S(1)=1))\slash \sum_{i=1}^n Z_iS_i$.
Next, $P^{11}(0)$ is estimated by first estimating $\alpha_0$ as the solution to
$\sum_{i=1}^n U^0(O_i;\alpha_0,\beta_0) = 0$ where
\begin{eqnarray}
U^0(O_i;\alpha_0,\beta_0) = Z_i\left( S_i - \widehat P(S(0)=1) 
\sum_{y=0}^1 w_0(y;\alpha_0,\beta_0)\widehat P(Y(0)=y|S(0)=1) \right), \label{eq: GBHform}
\end{eqnarray}

\noindent where $\widehat P(Y(0)=1|S(0)=1)$ is obtained in the same way as
 $\widehat P(Y(1)=1|S(1)=1)$.  Then with $\widehat \alpha_0$ from (\ref{eq: GBHform}),
$\widehat P^{11}(0) = [\widehat P(S(0)=1) \slash \widehat P(S(1)=1)] 
w_0(1;\widehat \alpha_0,\beta_0) \widehat P(Y(0)=1|S(0)=1)$. We implement this ``Standard SACE Method" verbatim multiple times below, with the definition of $S$ (and sometimes $Y$) changing for estimating needed terms in $CEP(s_1,s_0)$.

The Standard SACE Method requires $P(S(1)=1) < P(S(0)=1)$ in order that $\widehat P^{11}(z)$ for each $z=0,1$ has an asymptotic normal distribution and thus to ensure that Wald confidence intervals for $P^{11}(z)$ based on asymptotic or nonparametric bootstrap variance estimates have correct coverage probabilities (Jemiai et al., 2007).  Moreover, if $P(S(1)=1) < P(S(0)=1)$ but the probabilities are close, then the
Wald confidence intervals can have poor coverage.  Therefore, the Standard SACE Method only reliably gives correct inference if the data support $P(S(1)=1) < P(S(0)=1)$ with the probabilities not too close.  For implementation of this
method for inference on $CEP(s_1,s_0)$, the
needed SACE assumption for {\bf C} translated to the binary PSEM problem 
is A4$^{\prime \prime}$: $P(Y^{\tau}(1)=1) < P(Y^{\tau}(0)=1)$, whereas the needed SACE assumption for
${\bf A, B}$
is A5$^{\prime}$: $P(S^*(0)=1|Y^{\tau}=0) <  P(S^*(1)=1|Y^{\tau}=0)$.
Fortunately, both assumptions hold for many real binary PSEM data applications.  A5$^{\prime}$ essentially always holds, given that it is typically only interesting to study a binary marker as an effect modifier if it has a higher response rate in the active treatment arm 1 compared to the control arm 0.
A4$^{\prime \prime}$ typically holds in applications where 
the assumptions ${\bf C}$ hold, given that {\bf C} is motivated by applications 
where there is demonstrated or suspected treatment efficacy by time $\tau$, which is exactly A4$^{\prime \prime}$.
Moreover, A4$^{\prime \prime}$ and A5$^{\prime}$ are testable such that the conditions needed to assure valid inference can be checked.  In the simulations the required assumption 
A4$^{\prime \prime}$ or A5$^{\prime}$ holds, and in the example A5$^{\prime}$ holds whereas
A4$^{\prime \prime}$ is questionable.

{\it Standard SACE Method Accommodation of Inverse Probability Weighting.} The estimating equations for
$P(Y(1)=1|S(1)=1)$, $P(Y(0)=1|S(0)=1)$, $P(S(1)=1)$, $P(S(0)=1)$ and $\alpha_0$ are 
readily extended to include inverse probability weights in the standard way described above.

{\it Implementing the Standard SACE Method for $CEP(s_1,s_0)$.}
The semiparametric MLEs $\widehat{risk}_1(0,0)$ and 
$\widehat{risk}_0(0,0)$ are obtained as $\widehat P^{11}(1)$ and $\widehat P^{11}(0)$ described above, respectively, using $S \equiv [1 - Y^{\tau}][1-S^*]$ with {\bf B.0} and a fixed 
$\beta_0$.
The semiparametric MLEs $\widehat{risk}_1(1,1)$ and 
$\widehat{risk}_0(1,1)$ are obtained in the same way with $S \equiv [1 - Y^{\tau}]S^*$, with the wrinkle that the monotonicity assumption is in the reverse direction 
[i.e., $P(S(1) \ge S(0)) = 1$] from the published SACE method. 
This means that we use the Standard SACE Method reversing the roles of
$Z=1$ and $Z=0$, leading to a selection bias model 
$w_1(y;\alpha_1,\beta_1) \equiv P(S(0)=1|S(1)=1,Y(1)=y) = \lbrace 1 + {\rm exp}(-\alpha_1 - \beta_1 y) \rbrace^{-1}$ for $y=0,1$ with
sensitivity parameter $\beta_1$ defined by
${\rm exp}(\beta_1) = \frac{P(S(0)=1|S(1)=1,Y(1)=1) \slash \left\{ 1 - P(S(0)=1|S(1)=1,Y(1)=1) \right\} }{ P(S(0)=1|S(1)=1,Y(1)=0) \slash \left\{ 1 - P(S(0)=1|S(1)=1,Y(1)=0) \right\} }$.

The estimate $\widehat P^{11}(0)$ is obtained based
on $U^0(O_i;P(Y(0)=1|S(0)=1)) = (1-Z_i)S_i(Y_i - P(Y(0)=1|S(0)=1))\slash \sum_{i=1}^n (1-Z_i)S_i$ and $\widehat P^{11}(1)$ is obtained based on estimating $\alpha_1$ from $U^1(O_i;\alpha_1,\beta_1) = $
\begin{eqnarray}
(1-Z_i)\left( S_i - \widehat P(S(1)=1) 
\sum_{y=0}^1 w_1(y;\alpha_1,\beta_1)\widehat P(Y(1)=y|S(1)=1) \right), \label{eq: GBHform2}
\end{eqnarray}

\noindent 
and setting 
\begin{eqnarray}
\widehat P^{11}(1) = [\widehat P(S(1)=1) \slash \widehat P(S(0)=1)] 
w_1(1;\widehat \alpha_1,\beta_1) \widehat P(Y(1)=1|S(1)=1). \label{eq: GBHform2plus}
\end{eqnarray}

By standard estimating equation theory, the above estimators are consistent and asymptotically normal for given fixed $\beta_0$ and $\beta_1$. 
To obtain Wald confidence intervals for each $CEP(s_1,s_0)$, consistent estimating-function based variance estimators may be used 
for the estimates $\widehat P^{11}(z)$ not involving $\widehat \alpha_0$ or $\widehat \alpha_1$; e.g., the estimated variance of $\widehat P^{11}(1)$ for the Standard SACE Method
is given by 
$\sum_{i=1}^n \left( R_i \slash \pi(O_i,\widehat \psi)\right) \left[ U^1(O_i;\widehat P(Y(1)=1|S(1)=1)) \right]^2$.  Influence-function based variance estimates are similarly obtained for the estimates
$\widehat P^{11}(z)$ involving $\widehat \alpha_0$ or $\widehat \alpha_1$, by using a vector estimating function and the delta method. For example, for $\widehat P^{11}(1)$ estimated using equation (\ref{eq: GBHform2}), the four components of the estimating function are for 
$\widehat \theta = (\widehat \alpha_1, \widehat P(S(1)=1), \widehat P(S(0)=1), \widehat P(Y(1)=1|S(1)=1))^T$, with delta method applied with $g(w,x,y,z) = [x\slash y] w_1(1;w,\beta_1) z$.  All variance estimation is performed with the R package geex (Saul and Hudgens, 2017).\nocite{SaulHudgens2017}

To perform a sensitivity analysis, one approach specifies a plausible range $[l_k, u_k]$ (or maximum possible) for each sensitivity parameter $\beta_k$, $k=0,1$. An ignorance interval for $CEP(s_1,s_0)$ may be 
estimated by the span of values between 
the minimum and maximum estimates, obtained by setting $\beta_0$ and $\beta_1$ to the boundary values. Using the method of Imbens and Manski (2004)\nocite{ImbensManski2004} and
Vansteelandt et al. (2006)\nocite{Vansteelandtetal2006}, a Wald asymptotic (1-$\alpha$)\% estimated uncertainty interval (EUI) for $CEP(s_1,s_0)$ may be calculated 
as in formulas (40) and (41) of Richardson et al. (2014),\nocite{Richardsonetal2014}
using the variance estimates of the minimum and maximum $CEP(s_1,s_0)$ estimates. In particular, let
$\widehat{CEP}_l(s_1,s_0)$ and $\widehat{CEP}_u(s_1,s_0)$ be the estimates of $CEP(s_1,s_0)$ fixing the 
sensitivity parameters at the values within a pre-specified plausible region $\Gamma =
[l_0, u_0] \times [l_1,u_1]$ of the sensitivity parameter(s)
that minimize or maximize $\widehat{CEP}(s_1,s_0)$, respectively.  With $\widehat \sigma^2_l$ and $\widehat \sigma^2_u$ consistent estimates of the asymptotic limiting variances of $\widehat{CEP}_l(s_1,s_0)$ and $\widehat{CEP}_u(s_1,s_0)$, respectively, a (1-$\alpha$)\% EUI is given by
$[\widehat{CEP}_l(s_1,s_0) - c_{\alpha} \widehat \sigma_l \slash \sqrt{n},
\widehat{CEP}_u(s_1,s_0) + c_{\alpha} \widehat \sigma_u \slash \sqrt{n}]$, where $c_{\alpha}$ satisfies 
$$\Phi \left( c_{\alpha} + \left( \sqrt{n}(\widehat{CEP}_u(s_1,s_0)-\widehat{CEP}_l(s_1,s_0)) \right) \slash {\rm max} \lbrace 
\widehat \sigma_l, \widehat \sigma_u \rbrace  \right) - \Phi(-c_{\alpha}) = 1 - \alpha,$$

\noindent where $\Phi(\cdot)$ denotes the cdf of a standard normal variate. 
The same approach can be used to construct Wald confidence intervals and EUIs for the other scenarios and SACE approaches described below. Theoretical justification of these EUIs relies on the assumption that the values $\gamma_l, \gamma_u \in \Gamma$ that correspond to the 
ignorance interval for $CEP(s_1,s_0)$ are the same for all possible observed data laws (condition (39) from Richardson et al.), which holds in scenarios {\bf A} and {\bf B} and may need validation for scenario {\bf C} applications.

{\bf 4.3. Dichotomous Outcome SACE Methods Under Scenario {\bf B}}.

For scenario {\bf B}, $CEP(s_1,0)$ for $s_1=0,1$ can be estimated exactly as for scenario {\bf A}, with one change that
$SACE(1,1)$ vanishes because $p(1,1)=0$. In particular, first $\lbrace risk_1, risk_0\rbrace$ and $p(0,0)$ are 
estimated as in scenario {\bf A}, and then $p(1,0)$ is estimated as $\widehat p(1,0) = 1 - \widehat p(0,0)$. 
Secondly, 
$\lbrace risk_1(0,0),risk_0(0,0)\rbrace$ are estimated as in scenario {\bf A}.
Lastly, $risk_z(1,0)$ for each $z=0,1$ is estimated 
via equation (\ref{eq: riskz10}) plugging in estimates for each term. 
These steps amount to first estimating
$risk_1(0,0)$ by the solution to
$\sum_{i=1}^n Z_i(1-Y^{\tau}_i)(1-S^*_i)(Y_i - risk_1(0,0))\slash \pi(O_i,\widehat \psi) = 0$,
with convention that the summand is zero if $S^*_i=*$. Then $risk_0(0,0)$ and
$risk_0(1,0)$ are estimated by the solutions to the two equations {\bf B.0}
%
and $\widehat{risk}_0 - risk_0(0,0) \widehat p(0,0) - risk_0(1,0) \widehat p(1,0) = 0$; our code for the simulation study and example are implemented in this manner.

{\bf 4.4. Dichotomous Outcome SACE Methods Under Scenario {\bf C}}.

We implement a SACE method that relaxes monotonicity by using the sensitivity parameter $\beta_0$ in {\bf B.0} plus three additional sensitivity parameters:
\begin{align}
&{\bf B.2}\quad {\rm exp}(\beta_2)  =  \frac{risk_1(0,0) \slash \{1 - risk_1(0,0)\} }{ risk_1(0,*)\slash \{1 - risk_1(0,*)\} } \nonumber \\
&{\bf B.3} \quad{\rm exp}(\beta_3)  =  \frac{risk_1(1,0) \slash \{1 - risk_1(1,0)\} }{ risk_1(1,*)\slash \{1 - risk_1(1,*)\} } \nonumber \\
&{\bf B.4} \quad{\rm exp}(\beta_4)  =  \frac{ p(1,0) \slash \{1 - p(1,0)\} }{ P(S^*(1)=1|0,1) \slash \{1 - P(S^*(1)=1|0,1)\}}, \nonumber
\end{align}
\noindent where $risk_1(s_1,*) \equiv P(Y(1)=1|S^*(1)=s_1,S^*(0)=*,Y^{\tau}(1)=0,Y^{\tau}(0)=1)$ for $s_1=0,1$ and
$P(S^*(1)=1|0,1) \equiv P(S^*(1)=1|Y^{\tau}(1)=0,Y^{\tau}(0)=1)$.
The estimation steps are similar to those taken for Scenario {\bf B}, namely: First, $p(1,0)$ and $P(S^*(1)=1|Y^{\tau}(1)=0,Y^{\tau}(0)=1)$ are estimated as the solutions to the two equations {\bf B.4} and 
\begin{align}
&\widehat P(S^{*}(1)=1|Y^{\tau}(1)=0) - p(1,0) \widehat P(Y^{\tau}(0)=0|Y^{\tau}(1)=0)\nonumber \\ 
&\quad - P(S^*(1)=1|Y^{\tau}(1)=0,Y^{\tau}(0)=1) \{1- \widehat P(Y^{\tau}(0)=0|Y^{\tau}(1)=0)\} = 0.\nonumber
\end{align} 
Then $risk_0(0,0)$ and $risk_0(1,0)$ are estimated as in scenario {\bf B}. Finally, $risk_1(s,0)$ and $risk_1(s,*)$ are estimated as the solutions to  $\widehat P(Y(1)=1|Y^{\tau}(1)=0,S^{*}(1)=s) - $ \newline $risk_1(s,0) \widehat P(Y^{\tau}(0)=0,S^{*}(0)=0|Y^{\tau}(1)=0,S^{*}(1)=s) - risk_1(s,*)\{1- \widehat P(Y^{\tau}(0)=0,S^{*}(0)=0|Y^{\tau}(1)=0,S^{*}(1)=s)\} = 0$ and either {\bf B.2} for $s=0$ or {\bf B.3} for $s=1$.

\vspace{.14in}

\noindent {\bf 5. Simulation Study}

%
In the first simulation study, data were simulated under the assumptions of Scenario {\bf B} (A1--A4, Case CB). 
Each simulated data set contained $n$ independent individuals, with potential outcomes and observed random variables generated as follows.  First, $(Y^{\tau}(1),Y^{\tau}(0))$ was set to  (0,0) or (1,1) with probabilities 0.8 and 0.2, respectively. Thus A4 (EECR) holds. 
If $ Y^{\tau}(1) =1$, then $Y(0)$ and $Y(1)$ were set to 1. 
If $ Y^{\tau}(1) =0$, then $(S^{*}(1),S^{*}(0))$ was set to (0,0) or (1,0) with probabilities 0.4 and 0.6, respectively, such that Case CB holds.

To evaluate size and power of a test of $H_{0}: CEP(1,0) = CEP(0,0)$ versus $H_{1} : CEP(1,0) \neq CEP(0,0)$, data were simulated under 13 different assumed values for \newline $CEP(1,0) - CEP(0,0)$, namely $-0.6, -0.5, \ldots, 0.6$.
Specifically, 
if $Y^{\tau}(1)=0$ and $S^{*}(1) = S^{*}(0) = 0$, 
	then $Y(1)$ was generated as Bernoulli with mean $a$
	and $Y(0)$ as Bernoulli with mean 0.5.
On the other hand, 
if $Y^{\tau}(1)=0$ and $S^{*}(1) = 1, S^{*}(0) = 0$, 
	then $Y(1)$ was Bernoulli with mean $b$
	and $Y(0)$ was Bernoulli with mean 0.5.
Thus $CEP(1,0) - CEP(0,0) = b-a$ for contrast function $h(x,y)=x-y$.
The values of $a$ ranged from 0.7 to 0.1 by decrements of 0.05 and $b$ increased from 0.1 to 0.7 by increments of 0.05.  
   Under this parameterization $risk_0(0,0)=0.5=risk_0(1,0)$, implying $\beta_0$ = 0.

To generate the observed data, 
 $Z$ was drawn from a Bernoulli distribution with $P(Z = 1) = 0.5$, and the vector of observable random variables 
$(Z,Y^{\tau},S^*,Y)=(Z,Y^{\tau}(Z),S^*(Z),$ \newline $Y(Z))$ was determined. Simulations were conducted with and without case-cohort sampling of $S^*$.
For the latter,
membership in the random subcohort was determined by $R$, a Bernoulli random variable with mean $\nu$. 
For individuals with $Y^{\tau}=0$,
$S^*$ was observed for subcohort members (i.e., $R=1$)
and cases (i.e., $Y=1$).
Thus the observed random vector equaled
$(Z,Y^{\tau},Y,R,S^*I(Y^{\tau}=0, \max\{R,Y\}=1))$. 

Data sets were generated for all 351 combinations of: 
$n \in \{400,800,1600\}; \nu \in \{0.1,$ \newline $ 0.25, 1\}$;
and 
$CEP(1,0) - CEP(0,0) \in \{-0.6,-0.5,\ldots,0.6\}$.
For each of these combinations, 2000 data sets of size $n$ were generated.  
Analyses used $[l_{0}, u_{0}] \in \{[0,0],[-1,1],[-2.5,2.5]\}$.
For each simulated data set, $CEP(1,0) - CEP(0,0)$ was estimated under the  Scenario {\bf B} assumption set. 
The null hypothesis $H_0$ was rejected if and only if the 95\% EUI for $CEP(1,0)-CEP(0,0)$ excluded 0. 
Power was estimated by the proportion of simulated data sets where $H_0$ was rejected (Figure 1).  Empirical type I error was always less than or  approximately equal to the nominal significance level $\alpha=0.05$.
As expected, power increased with sample size and decreased as the interval $[l_0, u_0]$ became wider and as the size of the subcohort decreased.
Figure 2 shows the average widths of the 95\% EUIs for $CEP(1,0)-CEP(0,0)$.
The EUIs cover at approximately the nominal rate when $l_0=u_0$, i.e., when  $CEP(1,0)-CEP(0,0)$ is identifiable, and are conservative otherwise.
The widths are relatively constant across values of $CEP(1,0)-CEP(0,0)$ and increase as the size of the random subcohort decreases.
Empirical coverage of the 95\% EUIs are plotted in Web Figure 1 (Web Appendix C).
Web Figures 2 and 3 display bias 
of the $CEP(1,0)-CEP(0,0)$ estimates and ratios of the empirical standard errors (ESE) to the average of the sandwich variance estimated standard errors (ASE), showing unbiasedness of both the point and standard error estimators.

\begin{figure}
\centering
\includegraphics[scale=0.9]{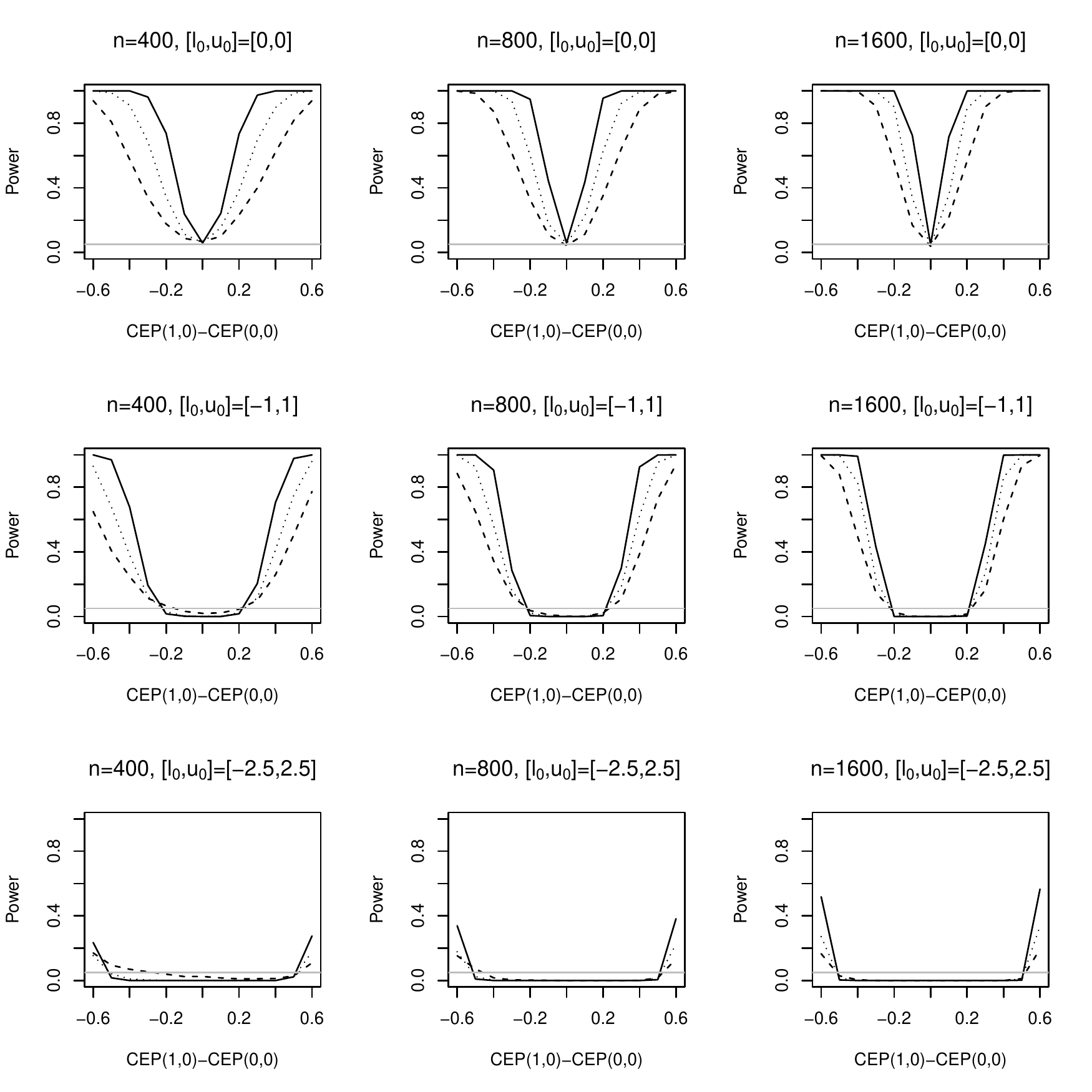}
\caption{Power to reject $H_{0}: CEP(1,0) = CEP(0,0)$ for the simulation study under Scenario {\bf B}.
Solid black lines denote full cohort and dashed (dotted) lines denote case-cohort with 10\% (25\%) random subcohort.
Horizontal gray lines denote significance level $0.05$.
}
\end{figure}

\begin{figure}
\centering
\includegraphics[scale=0.9]{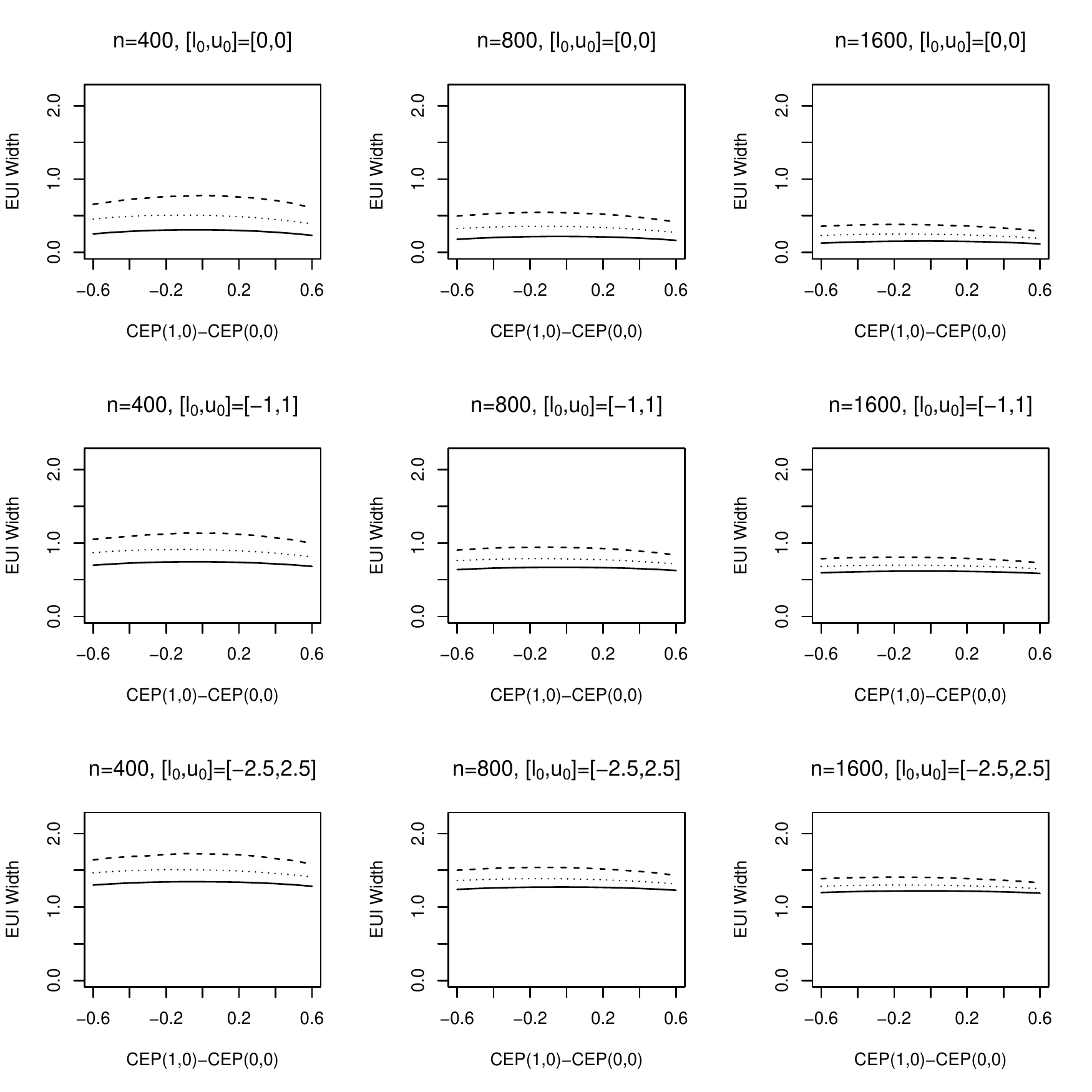}
\caption{Average 95\% EUI width for the simulation study under Scenario {\bf B}.
Solid lines denote full cohort and dashed (dotted) lines denote case-cohort with 10\% (25\%) random subcohort.
}
\end{figure}

The data sets simulated under Scenario {\bf B} were not analyzed with Scenario {\bf A} methods that allow $S^*(0)$ to vary, because, 
as described in the Introduction, in Scenario {\bf B} Case CB is known to be true structurally by the 
definition of $S^*(0)$.
In addition, it is not advisable to use Scenario {\bf C} methods because validity requires 
A4$^{\prime \prime}$: $P(Y^{\tau}(1)=1) < P(Y^{\tau}(0)=1)$ as discussed in Section 4.2, which can be diagnosed
to be dubious for many data sets because they are generated from a distribution 
with $P(Y^{\tau}(1)=1) = P(Y^{\tau}(0)=1)$. 

%
In the second simulation study, data were simulated under the assumptions of scenario {\bf C} (A1-A3, A4', Case CB) such that A4 in scenario {\bf B} failed. 
First $(Y^{\tau}(1),Y^{\tau}(0))$ was set to (0,0), (0,1), or (1,1) with probabilities 0.7, 0.2, and 0.1, such that A4' (ENHM) holds.
If $Y^{\tau}(z) = 1$, then $Y(z)$ was set to 1.
Similar to the first simulation, if $(Y^{\tau}(1),Y^{\tau}(0)) = (0,0)$, then $(S^{*}(1),S^{*}(0))$ was set to (0,0) or (1,0) with probabilities 0.4 and 0.6. If $(Y^{\tau}(1),Y^{\tau}(0)) = (0,1)$, then $S^{*}(1)$ was set to 0 or 1 with probabilities 0.4 and 0.6. 
 
To evaluate the power of the test of the same hypotheses as the first simulation, data were simulated under the same thirteen values of $CEP(1,0) - CEP(0,0)$ as before. 
%
Randomization assignment $Z$ and the observed data were also generated as in the first simulation,
for all 351 combinations of: $n \in \{2000,4000,8000\}; \nu \in \{0.1,0.25,1\};$ and $CEP(1,0) - $ \newline $CEP(0,0) \in \{-0.6, -0.5,\ldots ,0.6\}$. 
Analyses used $[l_{j},u_{j}] \in \left\{ [0,0],[-0.5,0.5],[-1,1] \right\}$ for $j=0,2,3,4$. 
As for the Scenario {\bf B} simulations, all data sets were simulated under no selection bias.  Results based on 2000 simulated data sets are shown in Figures 3 -- 4 and Web Figure 4 -- 6. As for scenario {\bf B}, power and precision diminished as the interval $[l_0, u_0]$ became wider and the subcohort size decreased.

\begin{figure}
\centering
\includegraphics[scale=0.9]{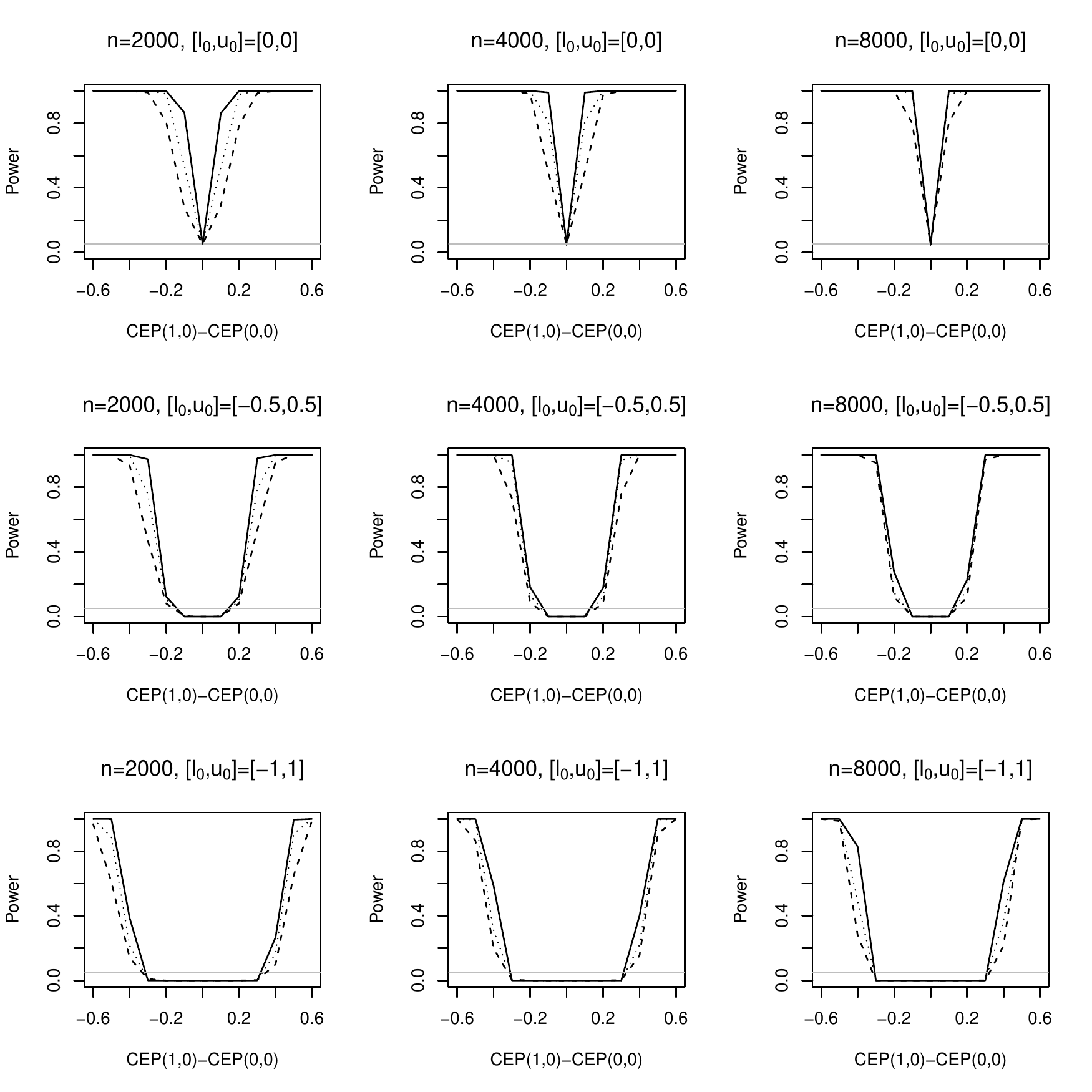}
\caption{Power to reject $H_{0}: CEP(1,0) = CEP(0,0)$ for the simulation study under Scenario {\bf C}.
Solid black lines denote full cohort and dashed (dotted) lines denote case-cohort with 10\% (25\%) random subcohort.
Horizontal gray lines denote significance level $0.05$.
}
\end{figure}

\newpage
\begin{figure}
\centering
\includegraphics[scale=0.9]{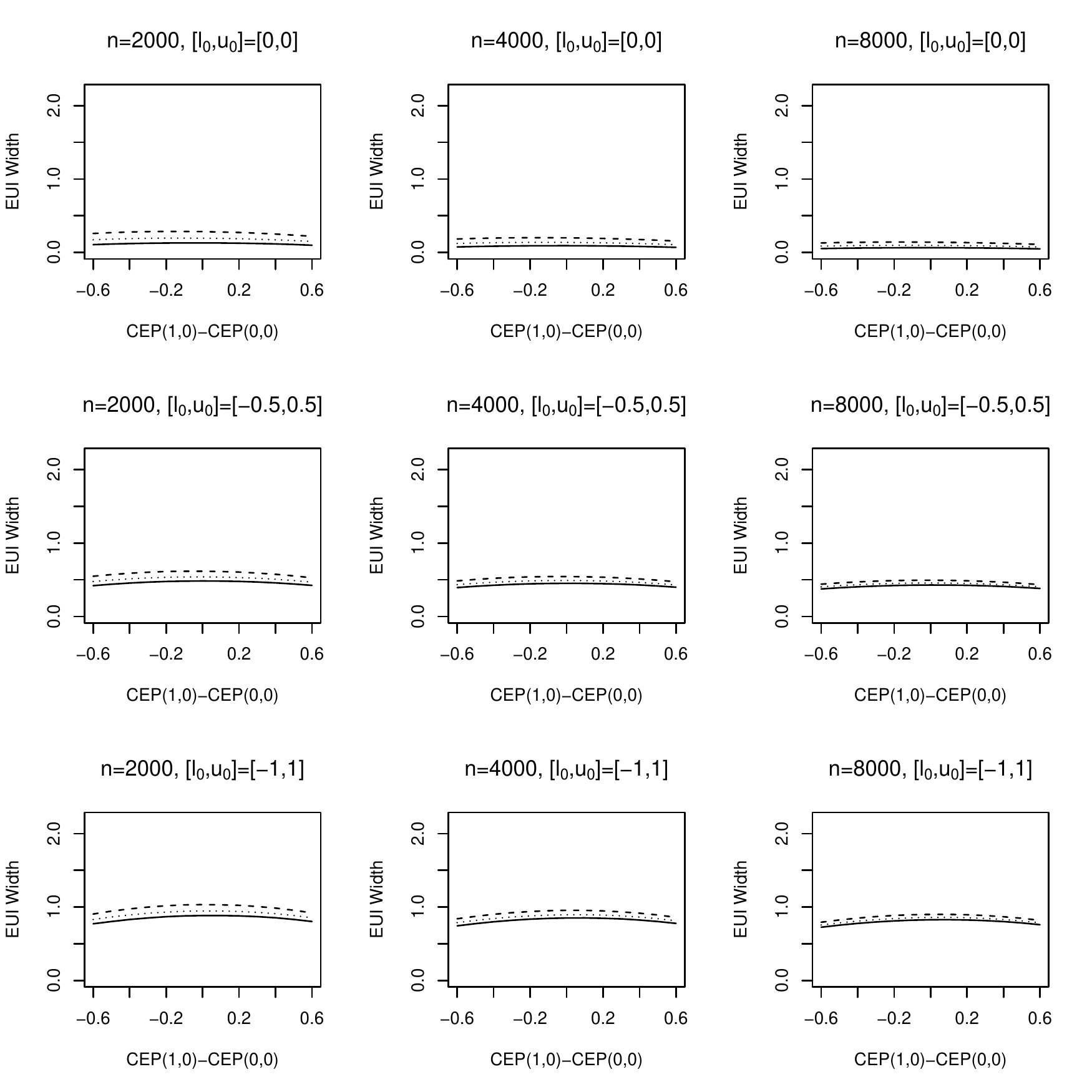}
\caption{Average 95\% EUI width for the simulation study under Scenario {\bf C}.
Solid lines denote full cohort and dashed (dotted) lines denote case-cohort with 10\% (25\%) random subcohort.
}
\end{figure}

\vspace{.14in}

\noindent {\bf 6. Application: HVTN 505 HIV Vaccine Efficacy Trial}

We apply the methods to HVTN 505 for three different binary markers defined based on the PFS or antibody measurements at Month 6.5 described in the Introduction.  These binary markers $S$ are the indicators of whether (S1) the PFS exceeds the median, (S2) the antibodies exceed the median, or (S3) either PFS or antibodies exceed the median. 
For marker S1, 70 of 125 (56\%) uninfected vaccinees have $S^*=1$, compared to 5 of 25 (20\%) infected vaccinees.  For S2 these rates are 66 of 125 (53\%) and 9 of 25 (36\%), respectively,
whereas for S3 these rates are 91 of 125 (73\%) and 10 of 25 (40\%).

Case CB holds in HVTN 505, as it generally does in preventive HIV vaccine efficacy trials, because biomarker values (which are measures of HIV-specific responses to vaccination) are zero for those assigned placebo.  
Given Case CB holds, we conduct the analysis under assumption scenarios {\bf B} and {\bf C}, 
and thus our goal is inference on $CEP(0,0)$ and $CEP(1,0)$. We use a vaccine efficacy contrast $h(x,y)=1-x\slash y$ such that the analysis assesses
$VE(0) \equiv CEP(0,0)$ and $VE(1) \equiv CEP(1,0)$, respectively.
The scenario {\bf B} method requires A5, which trivially holds, and A4 (EECR), which is defensible given that 
$\widehat P(Y^{\tau}(1)=1) = 14/1251 = 0.011$ and $\widehat P(Y^{\tau}(0)=1) = 10/1245 = 0.0080$ with Fisher's exact test 2-sided p-value 
of $p=0.54$ for a difference. In contrast, the scenario {\bf C} method requires A4$^{\prime}$, which relaxes A4 but at the cost of adding more sensitivity parameters. 
In addition, for Wald CIs and EUIs to have nominal coverage under scenario {\bf C},
A4$^{\prime \prime}$ ($P(Y^{\tau}(1)=1) < P(Y^{\tau}(0)=1)$) must hold as described in Section 4.2, which does not seem
to be true in this example.  Thus scenario {\bf B} may be the most reasonable, and we focus on it, but we also apply the method under scenario {\bf C} for illustration purposes.
The methods are implemented exactly as in the simulation study.  Web Appendix D studies why {\bf B} and {\bf C} are essentially equivalent for HVTN 505, where extra simulations suggest that a key part of the explanation is that
$P(Y^{\tau}(1)=1,Y^{\tau}(0)=0)$ is small.

Figure 5 shows the results in terms of ignorance intervals and 95\% EUIs under each of the three ranges of sensitivity parameters specified in the simulations, the top panel for $VE(1) - VE(0)$ and the bottom panel for $VE(0)$ and $VE(1)$. 
First, note that the scenario {\bf B} and {\bf C} methods give very similar results (top panel), with EUIs for the latter method only very slightly wider (and thus the bottom panel only shows results from the scenario {\bf B} method).   Second, 
for the antibody marker S2, there is no evidence for $VE$ modification, with 95\% EUIs about $VE(1) - VE(0)$ always including 0.
Third, the results are similar for markers S1 and S3, where if there is assumed
to be no selection bias (left panel), then there is clear evidence for $VE$ modification (with 95\% EUIs for $VE(1) - VE(0)$ comfortably excluding 0), whereas accounting for selection bias makes this result borderline significant with the EUIs just bordering 0 and depending on the amount of selection bias assumed (middle and right panels). For the PFS marker S2, $VE(1)$ is estimated to be 0.784, 0.723--0.821, and 0.624--0.844 under the models allowing no, intermediate, and highest degrees of selection bias, with 95\% EUIs 0.572--0.996, 0.483--0.975, and 0.304--0.973, respectively.  In addition, the point estimates of $VE(0)$ are consistently negative, suggesting a qualitative interaction may have occurred, although the 95\% EUIs always include 0 (for all three markers), suggesting no clear evidence of negative vaccine efficacy for vaccine recipients with low immune response.  Overall the results suggest that the vaccine may have conferred some protection for vaccine recipients with PFS responses exceeding the median, where another efficacy trial would be needed to verify this result.  The HIV vaccine field has ``moved on" from the DNA/rAd5 type of HIV vaccine platform, no longer considering it.  Thus these new results are significant for sounding a note of caution to completely giving up on this approach, in suggesting that if a new version of the regimen could be invented that induces high PFS responses in a vastly larger subgroup of vaccine recipients, then it could potentially confer high enough overall vaccine efficacy to confer worthwhile public health benefit. 

\begin{figure}
\centering
\includegraphics[scale=0.9]{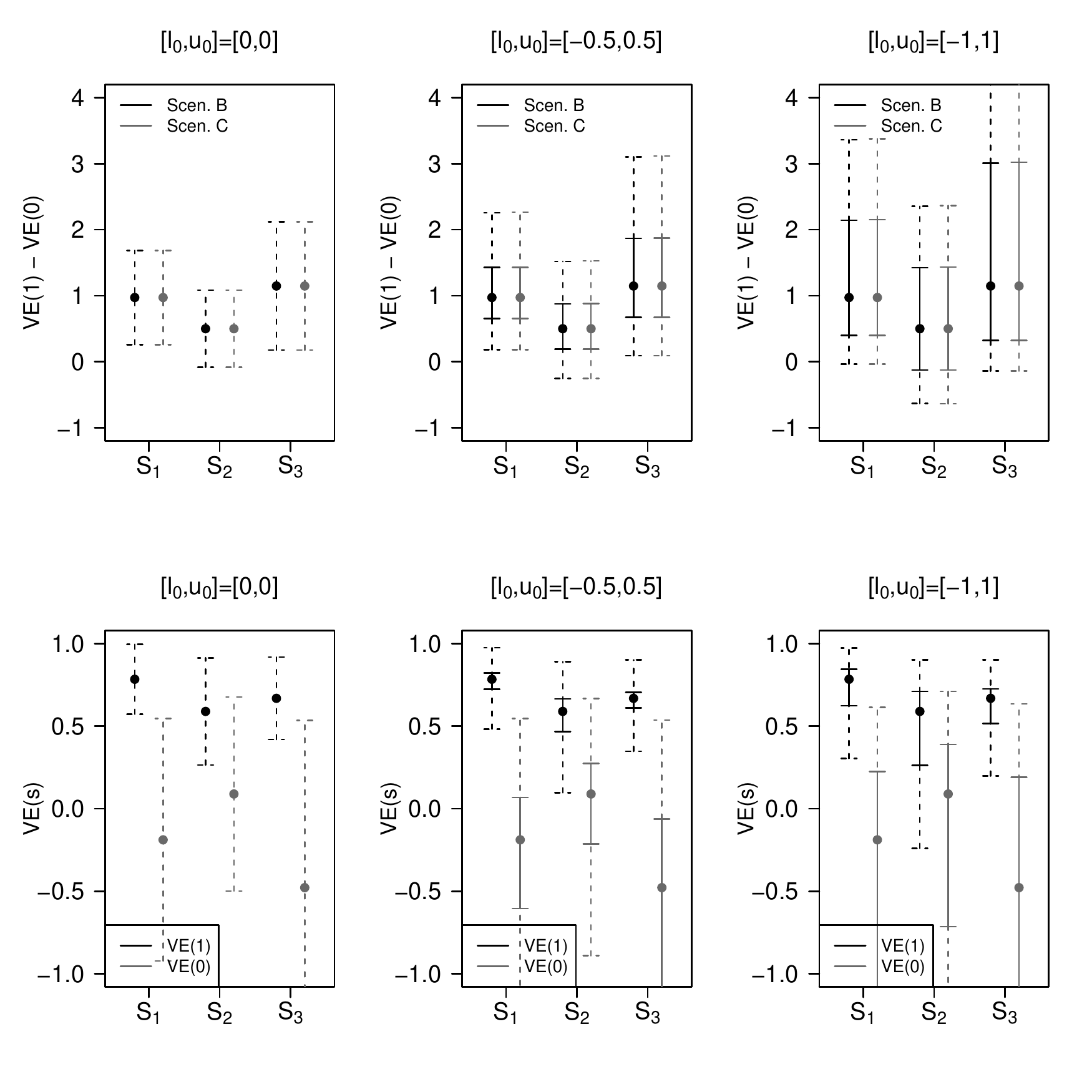}
\caption{For HVTN 505, ignorance intervals (solid lines) and 95\% EUIs (dashed lines) for $VE(0)=CEP(0,0)$ and $VE(1)=CEP(1,0)$ and $VE(1)-VE(0)$ with binary Month 6.5 marker $S^*$ defined by the indicator of whether (S1) the PFS exceeds the median, (S2) the antibodies exceed the median, or (S3) either PFS or antibodies exceed the median. 
The sensitivity analysis allows $\beta_0$ and $\beta_1$ to vary over $[l_0,u_0]=[0,0], [-0.5,0.5]$, or $[-1,1]$; circles are point estimates assuming no selection bias.  The top panel shows results for $VE(1)-VE(0)$ with the black and grey lines (left and right) results for the scenario {\bf B} and {\bf C} method, respectively. The bottom panel shows scenario {\bf B} results for $VE(1)$ and $VE(0)$.} 
\end{figure}

\vspace{.14in}

\noindent {\bf 7. Discussion}

A sizable literature on statistical methods for inference on the survival average causal effect (SACE) has developed over the past 20 years.  Motivated by a need for the HVTN 505 study, we described how these methods can be adapted to the problem of principal stratification effect modification (PSEM) evaluation in a randomized clinical trial with a binary intermediate response variable.  This provides new tools for PSEM analysis, expanding applicability to more settings, including: 
(1) to allow inference under a no negative early treatment effects assumption rather than the much stronger no early treatment effects assumption; (2) to allow inferences for studies with no adequate baseline predictors of the intermediate endpoint available and for which closeout placebo vaccination was not performed; and (3) to accommodate the sub-sampling design of the intermediate endpoint. These extensions were all 
needed for HVTN 505, with (1) relaxing a dubious assumption and (2) and (3) fitting the study design realities.  Moreover, the new methods avoid making a strong structural placebo conditional risk assumption made by most available PSEM methods, which was a necessary innovation for assessing a qualitative interaction in HVTN 505 given the appropriate skepticism that the DNA/rAd5 vaccine could have both beneficial and detrimental effects despite having no efficacy overall.
The data analysis provides the first direct evidence that vaccinated subgroups with high CD8+ T cell polyfunctionality score (PFS) responses to DNA/rAd5 vaccination had beneficial vaccine efficacy to prevent HIV-1 infection, which supports further research seeking to re-engineer the vaccine regimen to generate high PFS responses in more vaccine recipients.  


The new PSEM methods were developed under three common assumption scenarios that occur in randomized trials.  
An important decision for the analysis of a given data set is whether 
to use the scenario {\bf B} or {\bf C} method, with difference that the latter relaxes the
no early treatment effects assumption to monotonicity. For data sets with a small estimated value of $P(Y^{\tau}(1) = 1, Y^{\tau}(0) = 0)$, our simulations and application show that the 
methods tend to give similar results, with the intervals from scenario {\bf C} being only slightly wider than those from Scenario {\bf B}. Because there are many applications where monotonicity is much more plausible that no early treatment effects, in general the scenario {\bf C} method may be preferred unless there is compelling knowledge that no early treatment effects holds.



\vspace{.12in}
\noindent {\bf Supplementary Materials}
\begin{description}
\item[Title: Supportive results] The supportive results include (A) Application of Chiba and VanderWeele's (2011) SACE method for evaluating a binary principal stratification effect modifier under Scenarios {\bf A} and {\bf B};
(B) Adapting the SACE methods for a time-to-event outcome with right-censoring;
(C) Web Figures 1--6 simulation study results; 
(D) Additional Application results; and (E) Complete computer code.
\end{description}

\vspace{.12in}
\noindent {\bf Acknowledgements}

We thank the participants and investigators of HVTN 505, in particular Youyi Fong, Holly Janes, Georgia Tomaras, and Julie McElrath for providing the marker data for the example. Research reported in this publication was supported by the National Institute Of Allergy And Infectious
Diseases of the National Institutes of Health under Award Number R37AI054165 and 
the U.S. Public Health Service Grant AI068635.
The content is
solely the responsibility of the authors and does not necessarily represent the official views of
the National Institutes of Health.

\bibliography{ref}


\end{document}


\baselineskip=23.3pt
       
\nopageno                                                                                                                              
\begin{center}
{\Large {\bf
Web-based Supplementary Materials for: ``Post-randomization Biomarker Effect Modification in an HIV Vaccine Clinical Trial " by Peter B. Gilbert, Bryan Blette, Bryan E. Shepherd, and Michael G. Hudgens}}\\

\vspace{.25in}
Technical Report, Fred Hutchinson Cancer Research Center
May 23, 2018
\end{center}

\begin{center}
\end{center}

\section{Web Appendix A: Application of Chiba and VanderWeele's (2011) SACE Method for Evaluating a Binary Principal Surrogate Under Scenarios {\bf A} and {\bf B}}

We show how the simple SACE method of Chiba and VanderWeele (2011) can be applied to evaluate a binary principal surrogate using an additive contrast $h(x,y)=x-y$.
Define two sensitivity parameters $\alpha_k$, $k=0,1$, by
$\alpha_k \equiv P(Y(k)=1|Z=1,S_k=1) - P(Y(k)=1|Z=0,S_k=1)$ with
$S_0 \equiv [1-Y^{\tau}][1-S^*]$ and $S_1 \equiv [1-Y^{\tau}]S^*$.  Then
\begin{eqnarray}  
CEP(k,k;\alpha_k) & = & P(Y=1|Z=1,S_k=1) - P(Y=1|Z=0,S_k=1) - \alpha_k \label{eq: SACEkk} \\
          & = & \mu_{1k} - \mu_{0k} - \alpha_k \hspace{.2in} {\rm for} \hspace{.1in} k=0,1. \nonumber
\end{eqnarray}

\noindent A simple approach to estimating each 
$\mu_{zk}$, $(z,k) \in \lbrace (0,0), (0,1), (1,0), (1,1) \rbrace$, solves \newline
$\sum_{i=1}^n R_i U^{0zk}_i(O_i;\mu_{zk}) \slash \pi(O_i,\widehat \psi) = 0$ with
$U^{0zk}_i(O_i;\mu_{zk}) \equiv (1-Y^{\tau}_i)I(Z_i=z)I(S^*=k)(Y_i - \mu_{zk}) \slash n_{zk}$.
Then, $CEP(k,k;\alpha_k)$ is estimated by $\widehat \mu_{1k} - \widehat \mu_{0k} - \alpha_k$ where
$\alpha_k$ is a known constant fixed by the user. 
Lastly, plugging the above estimates into 
equation (2) [or equation (5)]
of the main article yields estimates of $risk_1(1,0;\alpha_1)$ and $risk_0(1,0;\alpha_0)$, and hence of $CEP(1,0;\alpha_0,\alpha_1)$. 

By standard estimating equation theory, the above estimators are consistent and asymptotically normal for given fixed $\alpha_0$ and $\alpha_1$. 
To obtain Wald confidence intervals for each $CEP(s_1,s_0;\alpha_0,\alpha_1)$, consistent sandwich variance estimators may be used,
e.g., the estimated variance of $\mu_{zk}$ for each $k=0,1$ is given by 
$\sum_{i=1}^n \left( R_i \slash \pi(O_i,\widehat \psi) \right) \left[  U^{0zk}_i(O_i;\widehat \mu_{zk}) 
\right]^2.$  The estimated variance of $\widehat{CEP}(1,0)$ may be obtained by the delta method.

To perform a sensitivity analysis, the user specifies a plausible range $[l_k, u_k]$ (or maximum possible) for each $\alpha_k$, $k=0,1$. An ignorance interval for $CEP(s_1,s_0)$ may be calculated 
as the minimum and maximum estimates (obtained with $\alpha_0$ and $\alpha_1$ set to the boundary values). Using the method of Imbens and Manski (2004)\nocite{ImbensManski2004} and
Vansteelandt et al. (2006)\nocite{Vansteelandtetal2006}, a Wald asymptotic (1-$\alpha$)\% estimated uncertainty interval (EUI) for $CEP(s_1,s_0)$ may be calculated 
as in formulas (40) and (41) of Richardson et al. (2014),\nocite{Richardsonetal2014}
using the variance estimates of the minimum and maximum $CEP(s_1,s_0)$ estimates. This approach requires that $CEP(k,k;\alpha_k)$ is monotone in $\alpha_k$, which holds by
(\ref{eq: SACEkk}).  The same approach can be used to construct Wald confidence intervals and EUIs for the other scenarios and SACE approaches described below.

\section{Web Appendix B: Adapting the SACE Methods for a Time-to-Event Outcome with Right-Censoring}

The approach to estimation of $CEP(0,0)$, $CEP(1,0)$, and $CEP(1,1)$ using the published SACE methods described above is similar if the binary outcome
$Y$ is defined as $Y \equiv I(T \le t)$ with $T$ subject to right-censoring and $t$ is a fixed time point of interest.
The estimating equations used to estimate the needed terms are 
the same as those described above, except that 
new estimating functions $U(O_i;\cdot)$ are swapped into the equations that are designed to handle
the right-censoring.  For example, consider the first estimating equation
in Section 3.2,
$\sum_{i=1}^n U^{0z}_i(O_i;risk_z)=0$.
With $Y \equiv I(T \le t)$, the same estimating equation can be used
swapping in the estimating function of the Kaplan-Meier estimator (Reid, 1981) or of the
targeted maximum likelihood estimator of a survival curve (Moore and van der Laan, 2009). The same type of swap is 
made for the other estimating equations.  As described in Shepherd et al. (2007, 2011), a modification to the weight functions $w_z(\cdot;\alpha_z,\beta_z)$ is needed,
where now they are indexed by time $t$:
$w_z(t;\alpha_z,\beta_z) = \lbrace 1 + {\rm exp}(-\alpha_z -\beta_z {\rm min}(t,\nu)) \rbrace^{-1}$, where $\nu$ is near the end of follow-up.
In addition, the summation $\sum_{y=0}^1 w_0(y;\alpha_0,\beta_0)\widehat P(Y(0)=y|S(0)=1)$ in equation (6) 
of the main article is changed to
$\int_0^{\infty} w_0(t;\alpha_0,\beta_0)\widehat P(T(0)\le t|S(0)=1)$ and 
the summation $\sum_{y=0}^1 w_1(y;\alpha_1,\beta_1)\widehat P(Y(1)=y|S(1)=1)$ in equation (7) 
of the main article is changed to
$\int_0^{\infty} w_1(t;\alpha_1,\beta_1)\widehat P(T(1)\le t|S(1)=1)$.

%
%
%

\newpage

\section{Web Appendix C: Additional Figures Showing Results of the Simulation Study}

\begin{center}
\begin{figure}
\includegraphics[scale=.85]{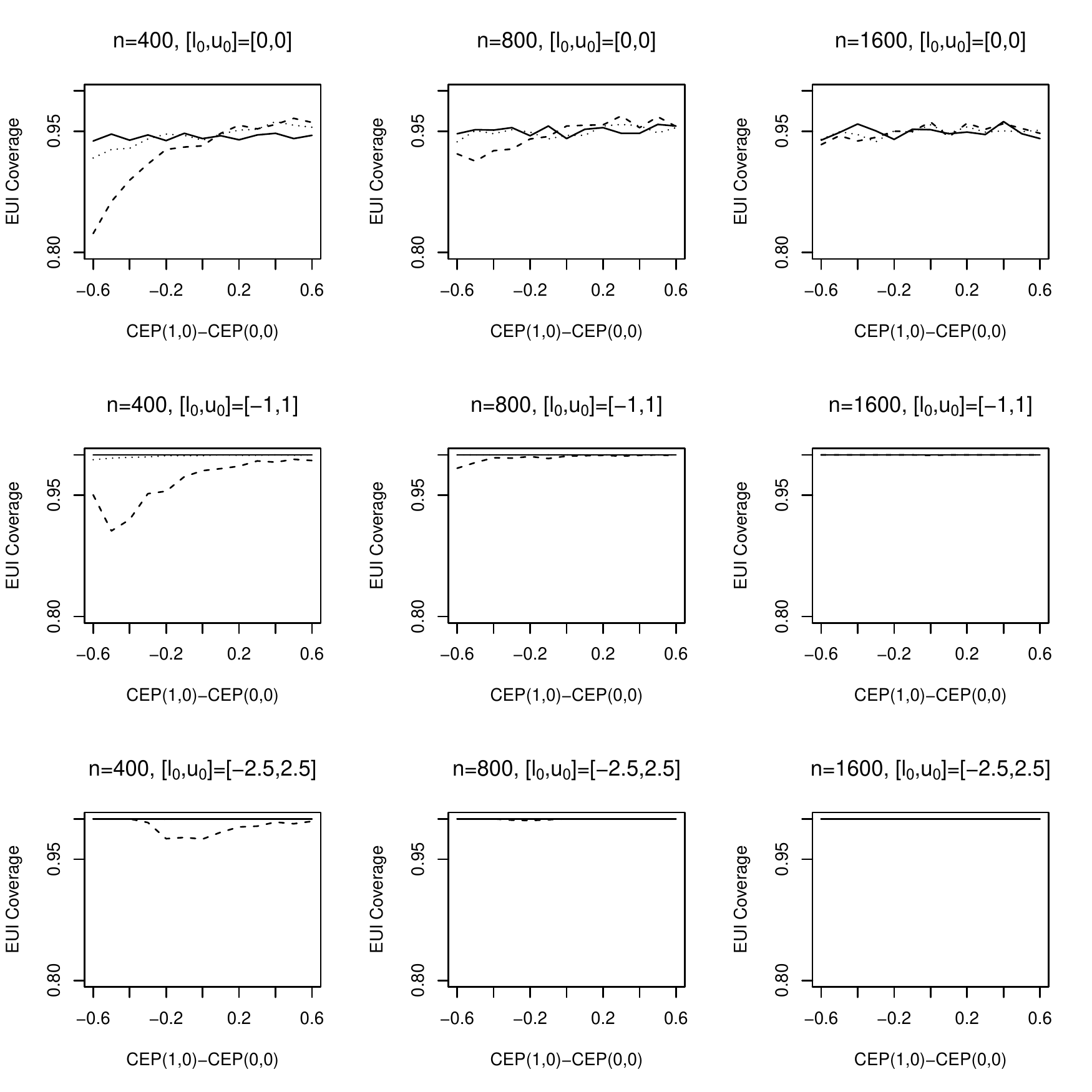}
\end{figure}
\end{center}
Web Figure 1. 95\% EUI coverage for simulation study under scenario {\bf B}.
Solid line denotes full cohort, dashed (dotted) line denotes case-cohort with 10\% (25\%) random subcohort.

\newpage
\begin{center}
\begin{figure}
\includegraphics[scale=.85]{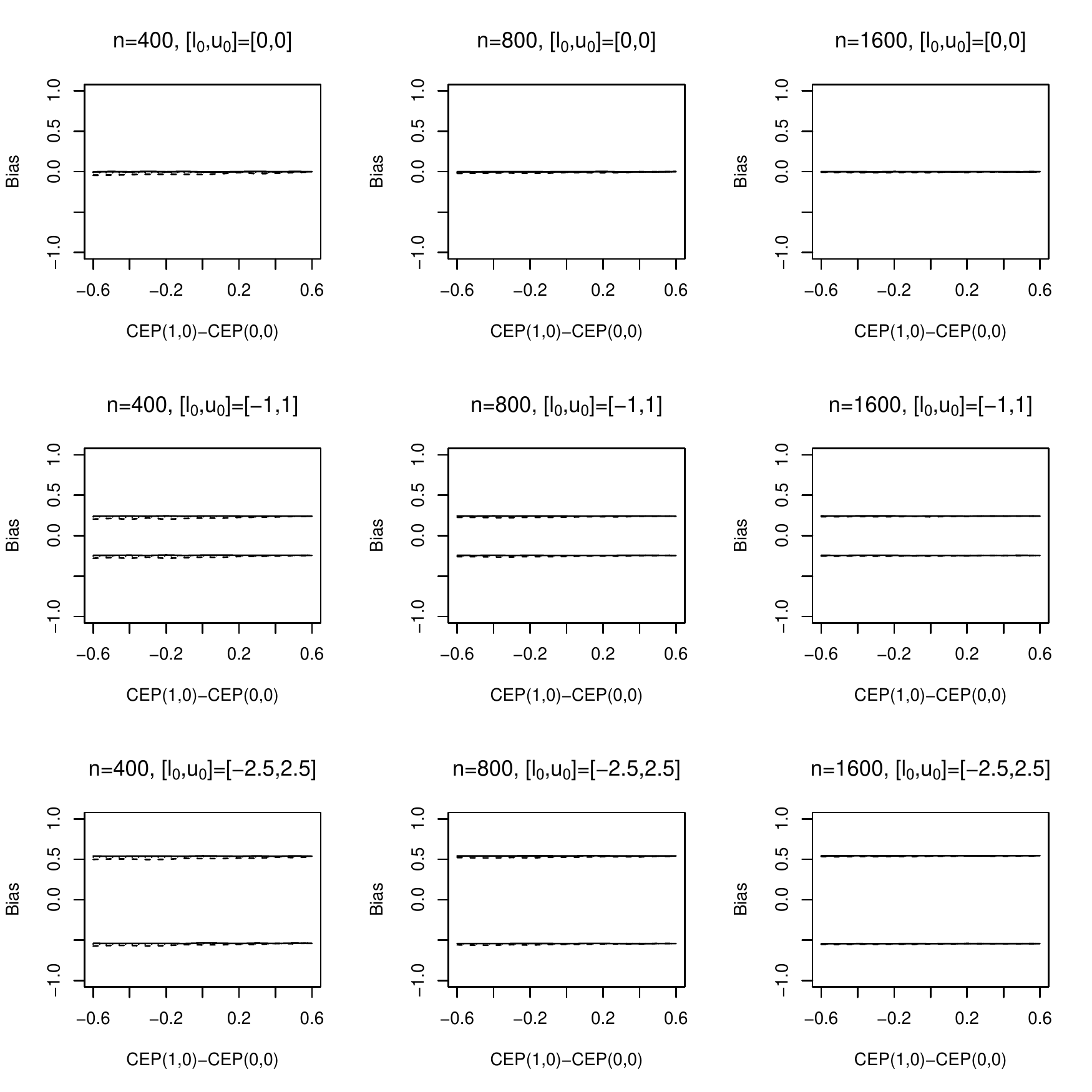}
\end{figure}
\end{center}
Web Figure 2. Bias for simulation study under scenario {\bf B}.
The middle and lower panels show bias of the minimum and maximum estimates of $CEP(1,0)-CEP(0,0)$ over
the plausible region $\Gamma=[l_0,u_0]$ of the sensitivity parameter $\beta_0$. 
Solid line denotes full cohort, dashed (dotted) line denotes case-cohort with 10\% (25\%) random subcohort.

\newpage
\begin{center}
\begin{figure}
\includegraphics[scale=.85]{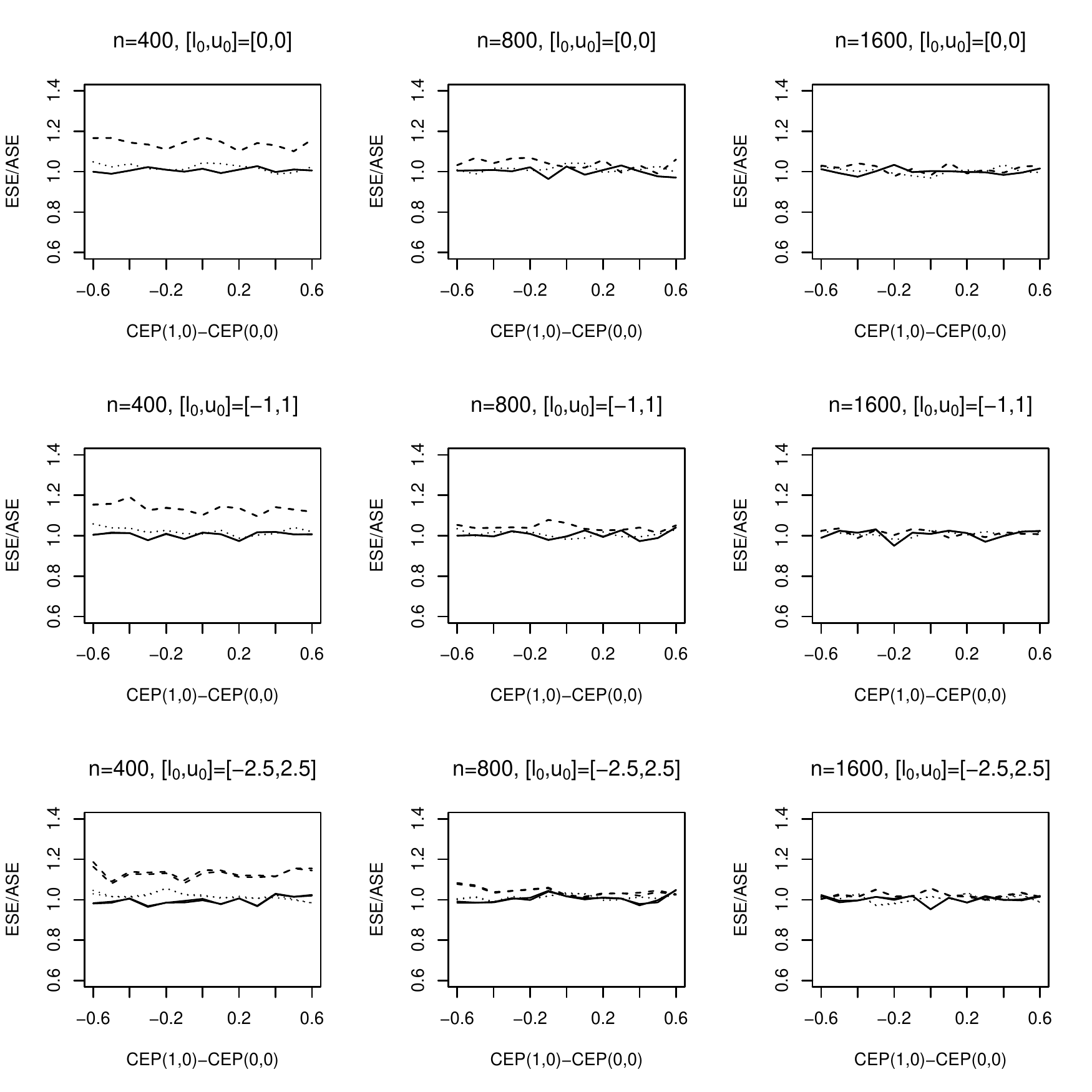}
\end{figure}
\end{center}
Web Figure 3.
Ratio of empirical standard error (ESE) to average estimated standard error (ASE) for simulation study under scenario {\bf B}.
Solid line denotes full cohort, dashed (dotted) line denotes case-cohort with 10\% (25\%) random subcohort.

\newpage
\begin{center}
\begin{figure}
\includegraphics[scale=.85]{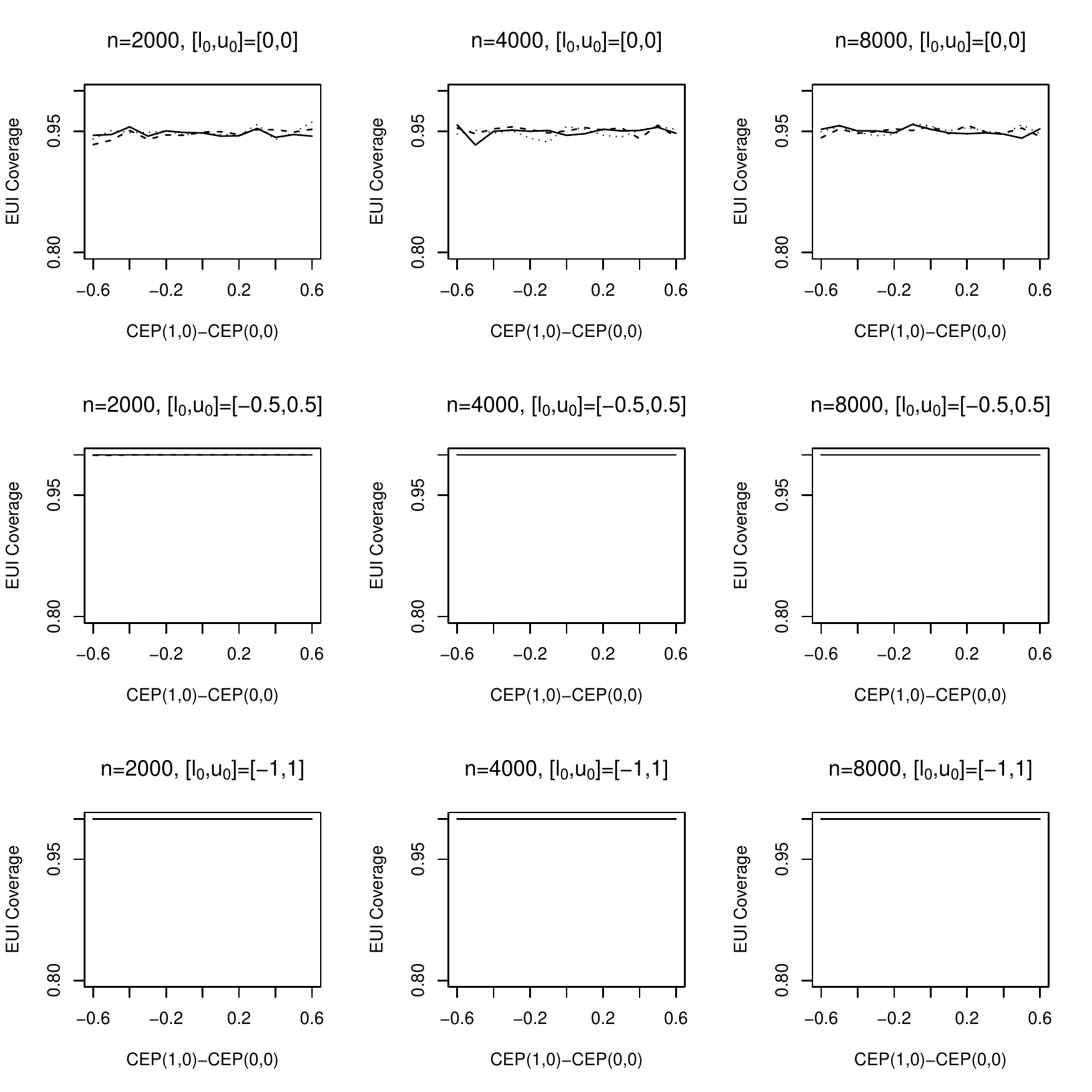}
\end{figure}
\end{center}
Web Figure 4.
95\% EUI coverage for simulation study under scenario {\bf C}.
Solid line denotes full cohort, dashed (dotted) line denotes case-cohort with 10\% (25\%) random subcohort.

\newpage
\begin{center}
\begin{figure}
\includegraphics[scale=.85]{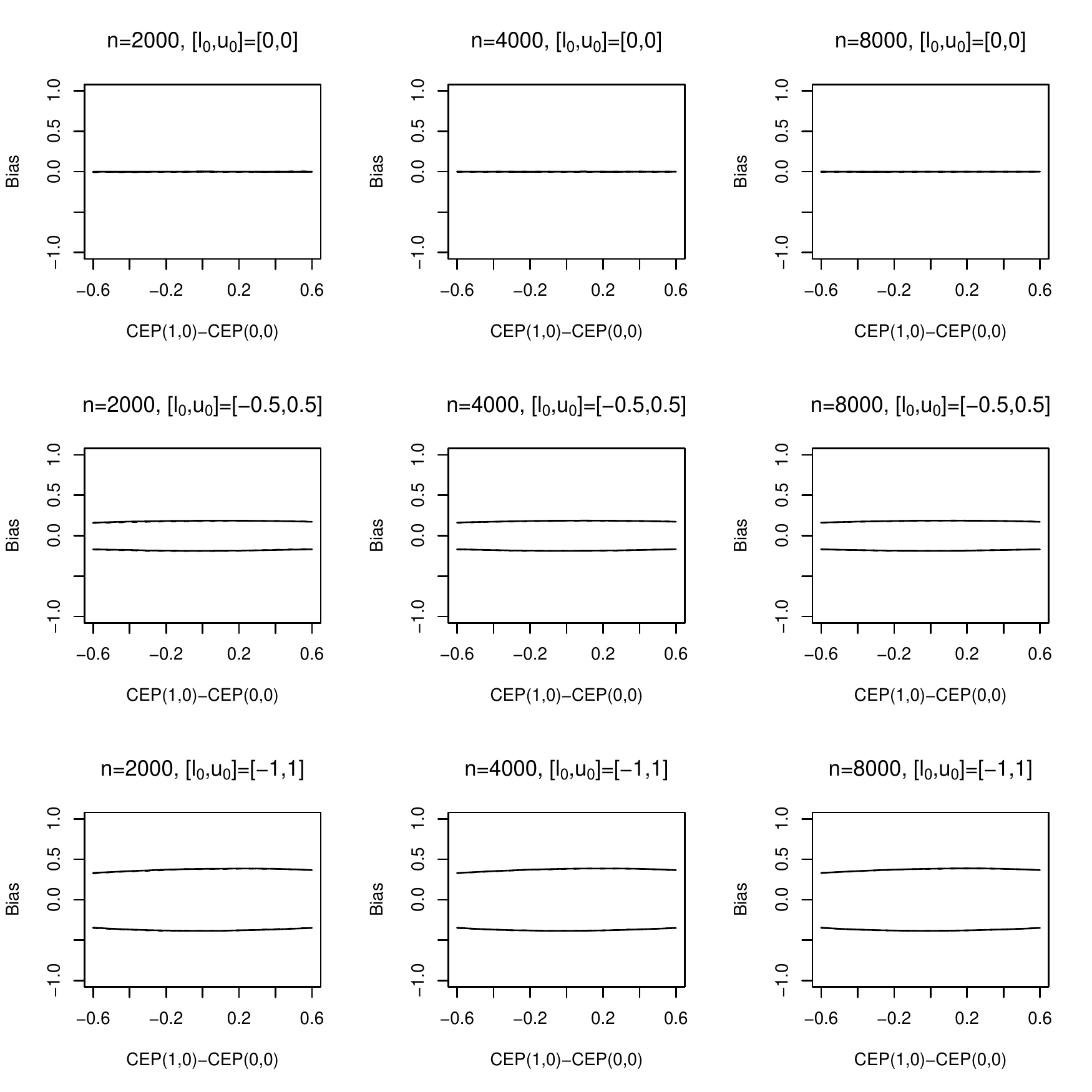}
\end{figure}
\end{center}
Web Figure 5. Bias for simulation study under scenario {\bf C}.
The middle and lower panels show bias of the minimum and maximum estimates of $CEP(1,0)-CEP(0,0)$ over
the plausible region $\Gamma=[l_0,u_0]$ of the sensitivity parameter $\beta_0$. 
Solid line denotes full cohort, dashed (dotted) line denotes case-cohort with 10\% (25\%) random subcohort.

\newpage
\begin{center}
\begin{figure}
\includegraphics[scale=.85]{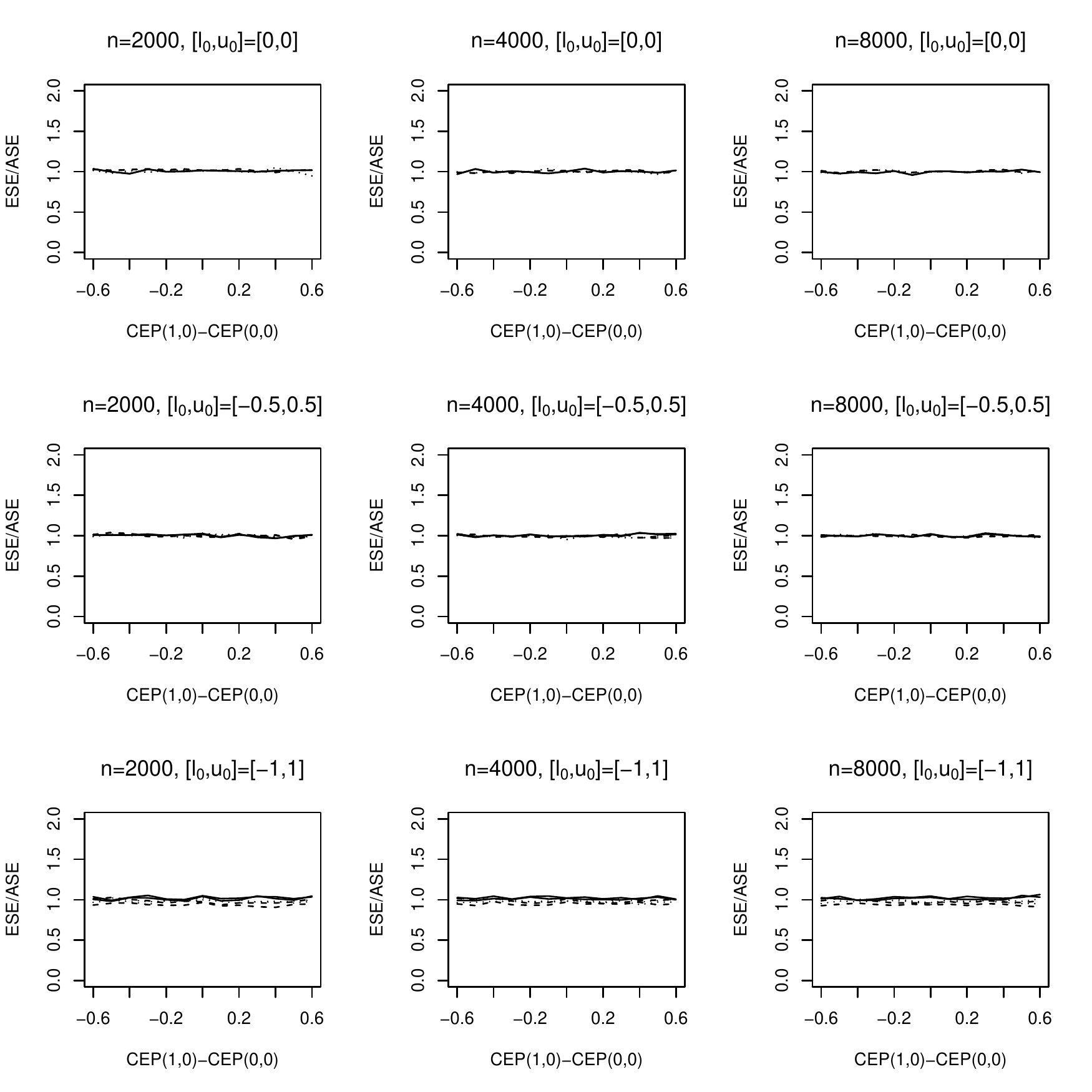}
\end{figure}
\end{center}
Web Figure 6. Ratio of empirical standard error (ESE) to average estimated standard error (ASE) for simulation study under scenario {\bf C}.
Solid line denotes full cohort, dashed (dotted) line denotes case-cohort with 10\% (25\%) random subcohort.

\newpage

\section{Web Appendix D: Additional Analyses for the Application}

\subsection{Explanation of How the Results of the Application are Very Similar for the Scenario {\bf B} and {\bf C} Methods}

In this web appendix we show that under the condition $P(Y(0)=1|Y^{\tau}(1)=1,Y^{\tau}(0)=0) =
P(Y(0)=1|Y^{\tau}(1)=0,Y^{\tau}(0)=0)$, then the scenario {\bf B} and scenario {\bf C}methods 
are equivalent.   
The typical estimand of interest in the article, $CEP(1,0)-CEP(0,0)$, is a function of $risk_{1}(1,0), risk_{1}(0,0), risk_{0}(1,0),$ and $risk_{0}(0,0)$. Under the setup for scenario {\bf C} in the application section, we switch the direction of the monotonicity assumption which, when combined with case CB, identifies $risk_{1}(1,0)$ and $risk_{1}(0,0)$. Thus, $risk_{1}(1,0)$ and $risk_{1}(0,0)$ are estimated in exactly the same way for the two scenarios:

\begin{align*}
\widehat{risk}_{1}(j,0) &= \widehat{P}(Y(1) = 1 | S^{*}(1) = j, S^{*}(0) = 0, Y^{\tau}(1) = 0, Y^{\tau}(0) = 0) \\
&= \widehat{P}(Y(1) = 1 | S^{*}(1) = j, Y^{\tau}(1) = 0) \\
&= \widehat{P}(Y = 1 | S^{*} = j, Y^{\tau} = 0, Z = 1), \quad j=0,1
\end{align*}

Therefore the results under each scenario will only differ in estimation of $risk_{0}(1,0)$ and $risk_{0}(0,0)$.

\noindent {\it Scenario B.}  Under scenario {\bf B}, $risk_{0}(1,0)$ and $risk_{0}(0,0)$ are estimated with a SACE method equivalent to solving the following mixing equations:

\begin{equation*}
exp(\beta_{0}) = \frac{risk_{0}(0,0) / (1-risk_{0}(0,0))}{risk_{0}(1,0) / (1-risk_{0}(1,0))}
\end{equation*}

\noindent and
\begin{equation*}
risk_{0} = p(0,0)risk_{0}(0,0) + (1-p(0,0))risk_{0}(1,0),
\end{equation*}

\noindent where $risk_{0}$ and $p(0,0)$ are identifiable under the assumptions of scenario {\bf B}. By EECR, we estimate:

\begin{align*}
\widehat{risk_{0}} &= \widehat{P}(Y(0) = 1 | Y^{\tau}(1) = 0, Y^{\tau}(0) = 0) \\
&= \widehat{P}(Y(0) = 1 | Y^{\tau}(0) = 0) = \widehat{P}(Y = 1 | Y^{\tau} = 0, Z = 0),
\end{align*}

\noindent which is solved by the estimating equation

\begin{equation*}
\sum_{i}(1-Z_{i})(1-Y^{\tau}_{i})(Y-risk_{0}) = 0.
\end{equation*}

In addition, by EECR and Case CB,

\begin{align*}
\widehat{p}(0,0) &= \widehat{P}(S^{*}(1) = 0, S^{*}(0) = 0 | Y^{\tau}(1) = 0, Y^{\tau}(0) = 0) \\
&= \widehat{P}(S^{*}(1) = 0 | Y^{\tau}(1) = 0) = \widehat{P}(S^{*} = 0 | Y^{\tau} = 0, Z = 1),
\end{align*}

which is solved by the estimating equation

\begin{equation*}
\sum_{i}Z_{i}(1-Y^{\tau}_{i})(1-S^{*}-p(0,0)) = 0.
\end{equation*}

\noindent Then $risk_{0}(1,0),$ and $risk_{0}(0,0)$ are estimated by solving the mixing equations once $\beta_{0}$ is specified.

\noindent {\it Scenario C.}
Under scenario {\bf C}, we relax the EECR assumption $[P(Y^{\tau}(1) = Y^{\tau}(0)) = 1]$ to 
ENHM $[P(Y^{\tau}(1) \geq Y^{\tau}(0)) = 1]$. Now, $risk_{0} = P(Y(0) = 1 | Y^{\tau}(1) = Y^{\tau}(0) = 0)$ is no longer identifiable. So first, an intermediate SACE method must be performed to estimate $risk_{0}$, and then the same SACE method as above is performed to estimate $risk_{0}(1,0)$ and $risk_{0}(0,0)$. We can write mixing equations in a similar form to those in the previous section (with some slight abuse of notation in defining $\beta_{1}$ as compared to the main text):

\begin{equation*}
exp(\beta_{1}) = \frac{risk_{0} / (1-risk_{0})}{P(Y(0) = 1 | Y^{\tau}(1) = 1, Y^{\tau}(0) = 0) / (1-P(Y(0) = 1 | Y^{\tau}(1) = 1, Y^{\tau}(0) = 0))}
\end{equation*}

\noindent and 
\begin{align*}
P(Y(0) = 1 | Y^{\tau}(0) = 0) &= P(Y^{\tau}(1) = 0 | Y^{\tau}(0) = 0)risk_{0} \\
&+ P(Y^{\tau}(1) = 1 | Y^{\tau}(0) = 0)P(Y(0) = 1 | Y^{\tau}(1) = 1, Y^{\tau}(0) = 0)
\end{align*}

\noindent where $P(Y^{\tau}(1) = 1 | Y^{\tau}(0) = 0) = 1 - P(Y^{\tau}(1) = 0 | Y^{\tau}(0) = 0)$.
Note that $\beta_{1}=0$ expresses the equality $P(Y(0)=1|Y^{\tau}(1)=1,Y^{\tau}(0)=0) =
P(Y(0)=1|Y^{\tau}(1)=0,Y^{\tau}(0)=0)$.  Now, when $\beta_{1} = 0$, we estimate

\begin{align*}
\widehat{risk}_{0} &= \widehat{P}(Y(0) = 1 | Y^{\tau}(1) = 1, Y^{\tau}(0) = 0) \\
&= \widehat{P}(Y(0) = 1 | Y^{\tau}(0) = 0) = \widehat{P}(Y = 1 | Y^{\tau} = 0, Z = 0),
\end{align*}

\noindent which is found by solving the same estimating equation as in scenario {\bf B}:

\begin{equation*}
\sum_{i}(1-Z_{i})(1-Y^{\tau}_{i})(Y-risk_{0}) = 0.
\end{equation*}

\noindent Therefore the estimate for $risk_{0}$ in scenario {\bf C} will be the same as that in scenario {\bf B} when $\beta_{1} = 0$. In addition, the estimate for $p(0,0)$ under scenario {\bf C}will be the same as that under scenario {\bf B} by monotonicity and Case CB:

\begin{align*}
\widehat{p}(0,0) &= \widehat{P}(S^{*}(1) = 0, S^{*}(0) = 0 | Y^{\tau}(1) = 0, Y^{\tau}(0) = 0) \\
&= \widehat{P}(S^{*}(1) = 0 | Y^{\tau}(1) = 0) = \widehat{P}(S^{*} = 0 | Y^{\tau} = 0, Z = 1),
\end{align*}

\noindent which is solved by the same estimating equation:

\begin{equation*}
\sum_{i}Z_{i}(1-Y^{\tau}_{i})(1-S^{*}-p(0,0)) = 0.
\end{equation*}

\noindent After estimating these terms, $risk_{0}(1,0)$ and $risk_{0}(0,0)$ are estimated via the same SACE method described for scenario {\bf B}. We have shown that when $\beta_{1} = 0$, the estimated $risk_{0}$ and $p(0,0)$ used in the mixing equations will be the same as those estimated in B. Thus, the estimated $risk_{0}(1,0)$ and $risk_{0}(0,0)$ (as well as the estimate of $CEP(1,0)-CEP(0,0)$) will be the same under both scenarios when $\beta_{1} = 0$.

As $\beta_{1}$ moves further away from 0, the estimates for $risk_{0}$ will become more different between the two methods, as will the subsequent results. Note that this is different from what was described in the methods and simulation sections, where the method for scenario {\bf C} is more complex due to the monotonicity assumption being in the usual direction.

\subsection{Repeating the Application with a new contrast function and a omparison of results in the Application for scenarios {\bf B} or {\bf C} under different early infection data distributions}

Web Figure 7 shows the identical results as Figure 5, except using a log relative risk contrast function $h(x,y) = $log$(x\slash y)$ and transforming the ignorance interval and EUI limits back to the $VE$ scale $h(x,y) = 1 - x\slash y$.

\begin{center}
\begin{figure}
\includegraphics[scale=0.9]{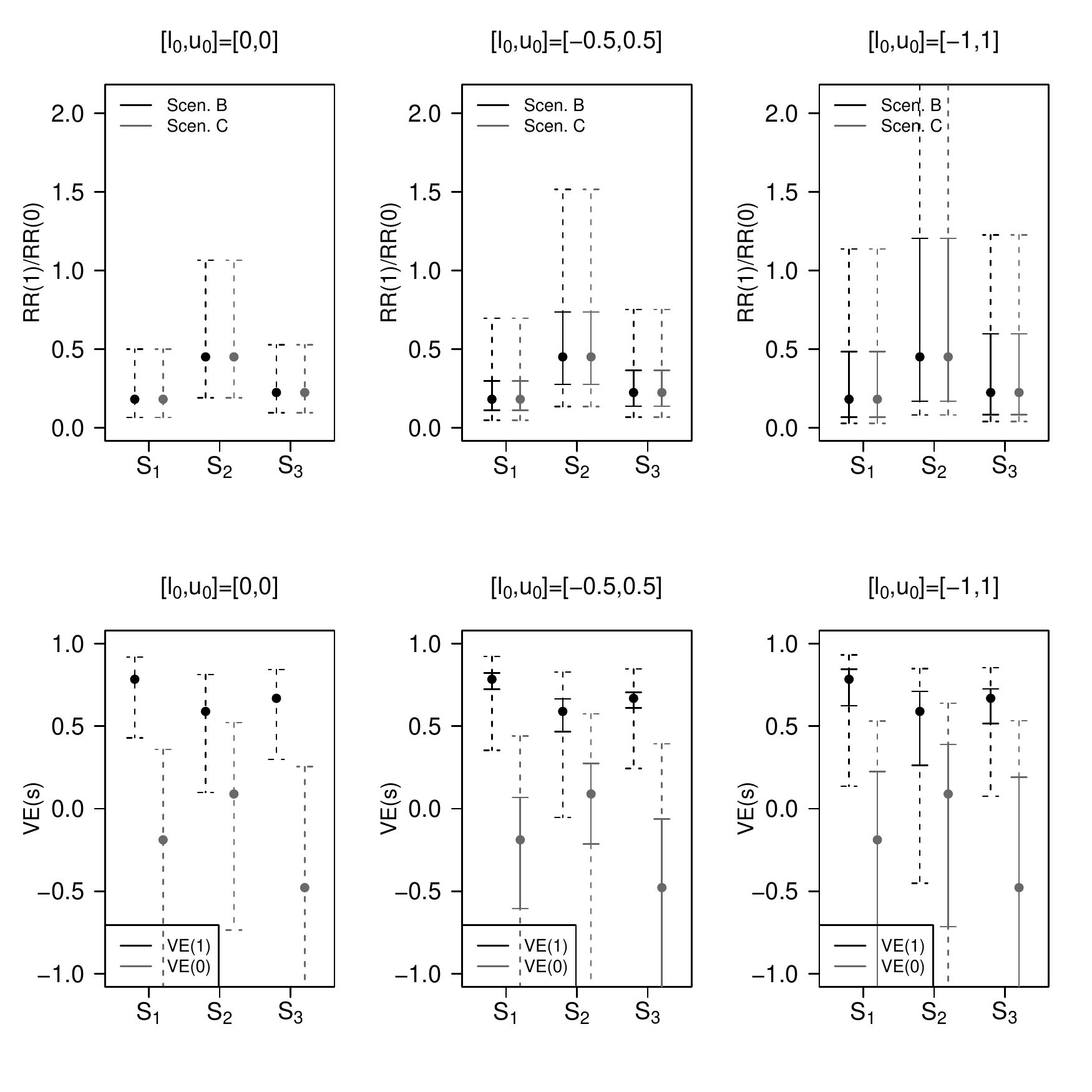}
\end{figure}
\end{center}
Web Figure 7.  For HVTN 505, ignorance intervals (solid lines) and 95\% EUIs (dashed lines) for $VE(0)=1-{\rm exp}(CEP(0,0))$ and $VE(1)=1-{\rm exp}(CEP(1,0))$ and $RR(1)/RR(0) = (1-VE(1))/(1-VE(0))$ with binary Month 6.5 marker $S^*$ defined by the indicator of whether (S1) the PFS exceeds the median, (S2) the antibodies exceed the median, or (S3) either PFS or antibodies exceed the median. 
The sensitivity analysis allows $\beta_0$ and $\beta_1$ to vary over $[l_0,u_0]=[0,0], [-0.5,0.5]$, or $[-1,1]$; circles are point estimates assuming no selection bias.  The top panel shows results for $RR(1)/RR(0)$ with the black and grey lines (left and right) results for the scenario {\bf B} and {\bf C} method, respectively. The bottom panel shows scenario {\bf B} results for $VE(1)$ and $VE(0)$.

As described in the application section of the main text, there may be situations where it is not clear whether the assumptions of scenario {\bf B} or C hold for the data being used. We study this issue by simulating the difference in EUI widths for the two scenarios under various supposed values of $P(Y^{\tau}(1) = 1, Y^{\tau}(0) = 0)$, which is implicitly assumed to be 0 under Scenario {\bf B}, but not under scenario {\bf C}. Each point on the plot in Web Figure 8 is the average of the differences in EUI widths calculated across 100 simulations. The red line represents the estimated value of $P(Y^{\tau}(1) = 1, Y^{\tau}(0) = 0)$ for the HVTN 505 data set analyzed in the application section.

\begin{center}
\begin{figure}
\includegraphics[scale=.5]{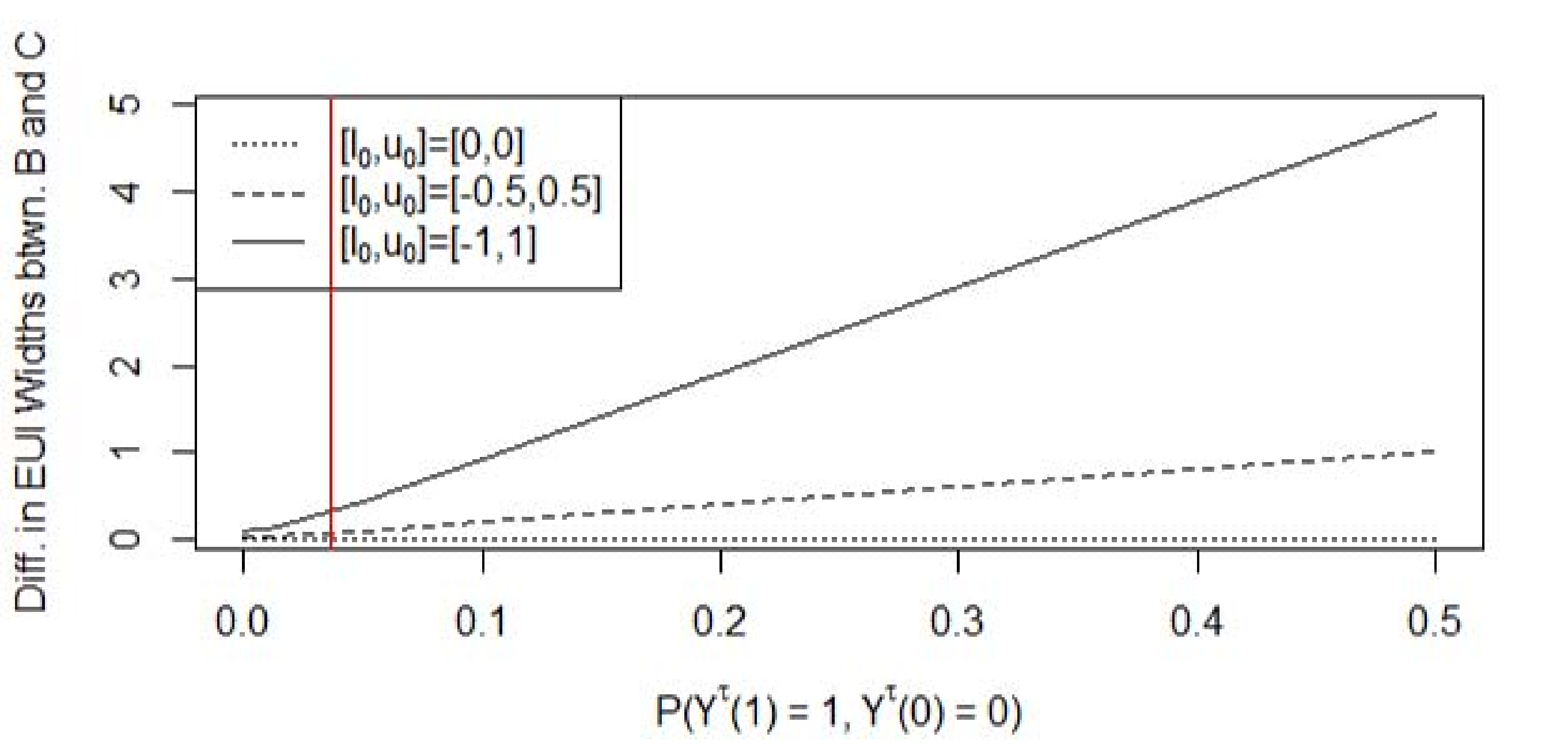}
\end{figure}
\end{center}
Web Figure 8. Difference in 95\% EUI width for the HVTN 505 application in the main article for analysis under scenario {\bf B} versus under scenario {\bf C}. 

For data sets with a small estimated value of $P(Y^{\tau}(1) = 1, Y^{\tau}(0) = 0)$, we would expect the two methods to give very similar results, with the intervals from scenario {\bf C} being only slightly wider than those from scenario {\bf B}. For such data sets it may be unclear whether to analyze under scenario {\bf B} or scenario {\bf C}. In general, justification for use of the scenario {\bf B} method would need to come from knowledge that the EECR assumption is plausibly true.  

\bibliography{ref}